\documentclass[twocolumn]{revtex4-2}
\usepackage[T1]{fontenc}
\usepackage[latin9]{inputenc}
\usepackage{float}
\usepackage{amsmath}
\usepackage{amssymb}
\usepackage{cancel}
\usepackage{graphicx}
\usepackage[colorlinks=false, pdfborder={0 0 0}]{hyperref}
\usepackage{braket}
\usepackage{cancel}
\usepackage[english]{babel}
\usepackage{algorithm}
\usepackage{algpseudocode}
\usepackage{tikz}
\usepackage{tikz-network}
\usepackage{adjustbox}
\usepackage{ragged2e}
\usepackage{subcaption}

\makeatletter

\renewcommand{\thetable}{\arabic{table}}
\renewcommand{\fnum@table}{Algorithm~\thetable}

\makeatother

\begin{document}
\title{Efficient Calculation of the Maximal R\'{e}nyi Divergence for a Matrix Product State via Generalized Eigenvalue Density Matrix Renormalization Group}
\author{Uri Levin, Noa Feldman, Moshe Goldstein }
\affiliation{Raymond and Beverly Sackler School of Physics and Astronomy, Tel-Aviv
University, Tel Aviv 6997801, Israel}

\begin{abstract}
The study of quantum and classical correlations between subsystems is 
fundamental to understanding many-body physics.
In quantum information theory, the quantum mutual information, $I(A;B)$, 
is a measure of correlation between the subsystems $A,B$ in a quantum state,
and is defined by the means of the von Neumann entropy: 
$I\left(A;B\right)=S\left(\rho_{A}\right)+S\left(\rho_{B}\right)-S\left(\rho_{AB}\right)$.
However, such a computation requires an exponential amount of resources. 
This is a defining feature of quantum systems, the infamous ``curse of dimensionality'' . 
Other measures, which are based on R\'{e}nyi divergences instead of
von Neumann entropy, were suggested as alternatives in a recent paper showing 
them to possess important theoretical features, and making them 
leading candidates as mutual information measures. 
In this work, we concentrate on the maximal R\'{e}nyi divergence. 
This measure can be shown to be the solution of a generalized eigenvalue problem. 
To calculate it efficiently for a 1D state represented as a matrix product state, 
we develop a generalized eigenvalue version of the density matrix renormalization group algorithm.
We benchmark our method for the paradigmatic XXZ chain, and show that the maximal R\'enyi divergence may 
exhibit different trends than the von Neumann mutual information.
\end{abstract}

\maketitle

\section{Introduction}
In the study many-body physics, correlation measures are key for characterizing phases and behaviors. In quantum systems, such correlation measures often originate from quantum information theory, such as the mutual information of two subsystems in a mixed state. It characterizes the total amount of correlations, both classical and quantum, between the subsystems \cite{Groisman_2005}. 
Beyond its conceptual significance, mutual information serves as a versatile diagnostic tool: it detects phase transitions \cite{Wicks_2007}, reveals entanglement structures \cite{swingle_2010}, and quantifies entanglement scaling in many body systems \cite{Eisert_2010}. 
In many-body physics, the scaling behavior of mutual information, such as area laws \cite{Wolf_2008}, provides deep insight into the nature of quantum correlations and the utility of tensor network representations.

Despite its great theoretical importance, in practice the calculation of the mutual information is often infeasible.
The mutual information is defined using the von Neumann entropy, $S\left(\rho\right)=-\mathrm{Tr}\left(\rho\log\rho\right)$, of the density matrices of the subsystems in question. 
The calculation of this measure requires the diagonalization of the subsystems density matrices, 
whose dimensions are exponential in the system size, 
making the overall calculation exponential and therefore impractical. 
This is true even for a 1D system represented by a matrix product state (MPS), unless the subsystems add up to the total system.
The unreachability of the von Neumann entropy is often solved by replacing it with R\'{e}nyi entropies, defined in Sec. \ref{subsec:Renyi-divergence} below, which are a family of measures computed out of the density matrix moments.
R\'enyi entropies and the R\'enyi mutual information, which is the result of substituting von Neumann entropy with R\'enyi's in the mutual information definition, have been studied extensively in recent years and were shown to be efficiently calculable.
In conformal field theory they can be calculated using the replica trick \cite{Pasquale_2004, Calabrese_2009}, 
they can be calculated for free fermions \cite{Bernigau_2015}, for states represented by a Matrix Product Density Operator (MPDO)
\cite{Pirvu_2010}, and by quantum Monte Carlo methods \cite{Cirac_2010, Hastings_2010, Humeniuk_2012, Grover_2013}.
These measures have been shown to characterize important phenomena such as quantum and thermal phase transitions 
\cite{Alcaraz_2014, Jean_Marie_2014, Singh_2011}, and the correlations in many-body localization \cite{Banuls_2017}.
However, when replacing the von Neumann entropy in the definition of the mutual information by the R\'enyi entropy, the resulting measure lacks important properties, ultimately making these entropy measures an unfitting replacement. 
Specifically, these measures can increase under local operations and some of them can even become negative \cite{Kormos_2017}.

Recently \cite{Scalet_2021}, R\'enyi-like measures for the mutual information were studied, and were shown to posses desirable theoretical features, such as non-negativity, and to be efficiently calculable. 
However, the methods offered for their calculation were technically limited.

One such measures is the maximal R\'enyi divergence \cite{Renner_2008} (relative entropy), calculated in the limit $\alpha\to\infty$ of the R\'enyi parameter.
This quantity has been shown to satisfy the desired theoretical properties (see Sec. \ref{subsec:Renyi-divergence}) and was proposed as a promising candidate for a mutual information measure.
Importantly, it can be formulated as a generalized eigenvalue problem (GEP) \cite{Ghojogh_2023}, a structure that arises in a wide range of physical contexts.
Methods for efficiently solving the GEP have been presented in different contexts \cite{Polizzi_2009, Verstraete_2004}, however these are
unsuited for our purposes in this work.

In this paper, we develop numerical tools within the framework of MPSs \cite{Cirac_2021, Eisert_2013} for solving GEPs in 1D using a generalized version of the well established density matrix renormalization group (DMRG) algorithm \cite{White_1992}. 
This is done with the efficient calculation of mutual correlations in mind, but in fact these tools are more general, capable of solving other problems described by a GEP.
These methods are shown to be more efficient than the approaches suggested by Ref. \citep{Scalet_2021}. 
We implement these tools in Python, using the TeNPy framework \cite{tenpy_1, tenpy_2} for the efficient manipulation of tensors.
Our implementation is available in a GitHub repository \cite{Levin_2025}.

The remainder of this paper is organized as follows. Sec. \ref{sec:background} provides the theoretical background underlying our work. 
Sec. \ref{subsec:MPS} introduces the MPS ansatz for representing one-dimensional systems, and Sec. \ref{subsec:The-DMRG-algorithm} outlines the DMRG method for determining ground states of one-dimensional operators. 
The Lanczos algorithm, which forms the computational core of DMRG, is described in Sec. \ref{subsec:Lanczos_Algorithm}. 
Sec. \ref{subsec:GEP} formulates the GEP, and Sec. \ref{subsec:Renyi-divergence} defines the R\'{e}nyi entropies and divergences.
Sec. \ref{sec:System-Edge-Calculation} then analyzes the maximal R\'{e}nyi divergence for subsystems at the edges of the system,
and Sec. \ref{sec:general-case-calculation} extends the discussion to the general case. 
Within the latter, Sec. \ref{subsec:Solving-the-GEP} presents the necessary generalization of the Lanczos algorithm, Sec. \ref{subsec:Technical-Implementation-Details} details implementation considerations, and Sec. \ref{subsec:Application-to-DMRG} demonstrates their application within the DMRG framework.
Finally, Sec. \ref{sec:results} reports numerical results from a representative physical model, and Sec. \ref{sec:conclusions} concludes with a summary of our findings.
Appendix \ref{sec:Irrelevance-of-Kernel} shows in detail that while null vectors of the subsystem density matrices should not change the maximal R\'{e}nyi divergence, they might cause numerical issues which require regularization.
Appendix \ref{sec:Generalized-Lanczos-eq} presents our adaptation and technical implementation details of the generalized Lanczos algorithm.
Appendix \ref{sec:gen-lanczos-correctness} proves the correctness of the generalized Lanczos algorithm, and discusses its convergence rate.

\section{background}\label{sec:background}
\subsection{MPSs and the Orthogonality Center}\label{subsec:MPS}

\begin{figure*}[ht]
    \centering
    \begin{subfigure}[b]{0.3\textwidth}
        \centering
        \includegraphics[width=\linewidth]{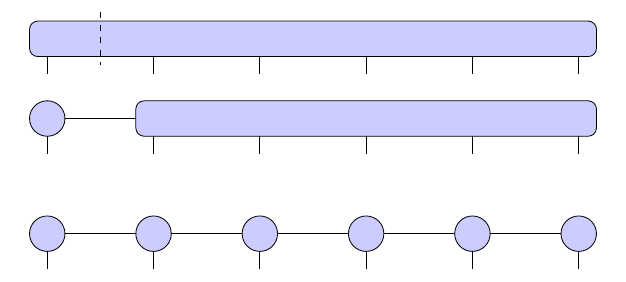}
        \caption{}
        \label{fig:MPS-construction}
    \end{subfigure}\hfill
    \begin{subfigure}[b]{0.3\textwidth}
        \centering
        \includegraphics[width=\linewidth]{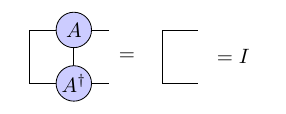}
        \caption{}
        \label{fig:Left-canonity}
    \end{subfigure}\hfill
    \raisebox{-0.05\height}{
        \begin{subfigure}[b]{0.3\textwidth}
            \centering
            \includegraphics[width=\linewidth]{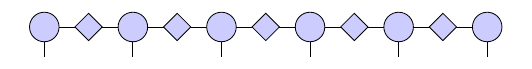}
            \caption{}
            \label{fig:MPS-Singular}
            \vspace{1em} 
            \includegraphics[width=\linewidth]{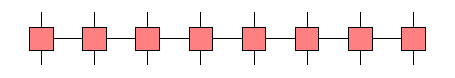}
            \caption{}
            \label{fig:MPO}
        \end{subfigure}
    }
    \label{fig:MPS-Intro}
    \caption{\justifying Tensor network graphical notations. (a) Construction of an MPS
    from a high-dimensional quantum state vector. The state can be written as: 
    $\ket{\Phi}=\sum_{\sigma_{1}\dots\sigma_{N}}\Phi_{\sigma_{1}\dots\sigma_{N}}\ket{\sigma_{1}\dots\sigma_{N}}=\sum_{j}A^{s_{1}\dots s_{i}}v^{i}A^{s_{i+1}\dots s_{N}}\ket{s_{1}\dots s_{N}}$,
    with $A^{s_{i}}$ site matrices of site $i$. Their contraction yields back the vector components $\Phi_i$,
    showing that the MPS does represent a state.
    (b) Left-orthogonality in an MPS. 
    Each tensor to the left of the orthogonality center satisfies
    the condition of Eq. \eqref{eq:left-canonical}, ensuring that contractions
    of these tensors with their conjugates yield the identity on the virtual
    bonds. This structure enables efficient and stable computation of
    observables and local optimizations.
    (c) Structure of an MPS with each set of Schmidt values explicitly depicted, representing Eq. \eqref{eq:MPS-definition-with-S}. 
    In this formulation, site tensors $\Gamma^{s_i}$ (circles) are not canonical on their own, 
    but can be made left (right) canonical by contracting with the Schmidt values $\Lambda^i$
    (diamonds) to their left (right).
    (d) Structure of an MPO. Each
    site tensor in the MPO carries two physical indices (input and output)
    and one or two virtual indices, allowing it to represent local operator
    actions and their correlations across sites. The MPO structure mirrors
    and complements the MPS ansatz, enabling efficient representation
    and application of many-body operators, such as Hamiltonians, within
    the tensor network framework.
    }
\end{figure*}

MPSs \cite{Schollwock_2011} are an important tool in the numerical
study of one-dimensional quantum many-body systems, especially within
the context of the DMRG \cite{White_1992} algorithm, detailed below. 
An MPS expresses the quantum state of a chain of sites as a product of site-dependent tensors, 
thereby enabling a compact representation of states obeying an area-law scaling of their entanglement.

Formally, a state $\ket{\Phi}$ of a system with $N$ sites may be
represented by an MPS in the following way. The MPS is composed of
a chain of rank 3 tensors $A$:
\begin{equation}
\label{eq:MPS-definition}
\ket{\Phi}=\sum_{\{s_{i}\}}A^{s_{1}}A^{s_{2}}\cdots A^{s_{N}}\ket{s_{1}s_{2}\dots s_{N}}.
\end{equation}

Each $A^{s_{i}}$ is a complex matrix and $s_{i}\in\left\{ 1\dots d\right\} $
runs over the local Hilbert space at the site $i$, where $d$ is
the local Hilbert space dimension. This representation is illustrated
in Fig. \ref{fig:MPS-construction}. We define the bond dimension
of the MPS, denoted by $\chi$, as the maximal inner dimension of the $A$ matrices. 
Using this notation, the size of each site tensor is at
most $d\chi^{2}$, and overall the size of the MPS is at most $Nd\chi^{2}$.
While in the general case, $\chi$ is required to be exponential in $N$, 
in many physically-interesting states $\chi$ may be polynomial in $N$, 
making the MPS an efficient ansatz. Additional computational 
efficiency can be achieved using truncation of the MPS bond dimensions.
Practically, this can be done by limiting the size or amount of singular values of each bond leg.
This allows for a low rank approximation of the state, 
which preserves relevant entanglement features \citep{Schollwock_2011}.

A critical concept in manipulating MPS is the \emph{orthogonality
center}, which refers to a specific site in the chain where the MPS
is locally normalized, and all tensors to the left (right) of it are
left- (right-) orthonormal. That is, for sites left of the orthogonality
center $k$, the tensors $A^{s_i}$ are left-canonical, satisfying:
\begin{equation}
\label{eq:left-canonical}
\sum_{\{s_{i}\}}\left(A^{s_{i}}\right)^{\dagger}A^{s_{i}}=I,
\end{equation}
while those on the right are right-canonical matrices, denoted by $B_{i}^{s_{i}}$, satisfying:
\begin{equation}
\label{eq:right-canonical}
\sum_{\{s_{i}\}}B^{s_{i}}\left(B^{s_{i}}\right)^{\dagger}=I.
\end{equation}
This canonical form simplifies many operations, including optimization
and expectation value calculations, by reducing the effective degrees
of freedom and ensuring numerical stability. Graphically, these properties
are shown in Fig. \ref{fig:Left-canonity}.

These canonical forms can be calculated from a state, by iteratively decomposing the state using the Singular Value Decomposition (SVD).
Starting from the left, as depicted in Fig. \ref{fig:MPS-construction}, the state is decomposed to a left canonical, unitary matrix $U$, singular values matrix $S$ and the remainder of the state $V^\dagger$.
The singular values are contracted to the right, leading to the next iteration.
A similar procedure can be applied beginning from the right, and at each step contracting the singular value to the remainder of the state, at the left.
These procedures decompose the state to a list of canonical tensors, and allow access to the Schmidt values at each site. 
While, in practice, the MPS is maintained as a list of canonical tensors, 
it is convenient to consider it decomposed to non-canonical site tensors and singular value matrices
at each bond, such that a contraction of the singular values to the left (right) will yield a
left (right) canonical tensor. In this form, Eq. \eqref{eq:MPS-definition} is rewritten as:
\begin{equation}
\label{eq:MPS-definition-with-S}
\ket{\Phi}=\sum_{\{s_{i}\}}
    \Gamma^{s_{1}} \Lambda^{1} \Gamma^{s_2} \cdots \Lambda^{N-1} \Gamma^{s_N}\ket{s_{1}s_{2}...s_{N}} ,
\end{equation}
as depicted in Fig. \ref{fig:MPS-Singular}, where the matrices $\Lambda^{i}$ are diagonal with the singular values on the diagonal.

A similar construction can be done for operators, which are called
\emph{Matrix Product Operators} (MPOs), seen in Fig. \ref{fig:MPO}.
Formally, an operator $O$ can be written as an MPO using:
\begin{equation}
\label{eq:MPO-definition}
O=\sum_{\mathbf{s,s'}}M^{s_{1}s'_{1}}M^{s_{2}s'_{2}}\cdots M^{s_{N}s'_{N}}\ket{\mathbf{s}}\bra{\mathbf{s'}}.
\end{equation}
Since operators are not generally normalized or constrained, MPOs
do not poses the orthogonal structure of MPS. Similarly to MPS, The
description of operators as MPOs allow for their efficient storage,
and it complements the MPS ansatz and allow for efficient calculations.

Another closely related structure is the iMPS (infinite MPS), which acts as an efficient representation of the thermodynamic limit of an MPS \cite{Vidal_2007}.
In contrast to a finite MPS, which uses an open or periodic boundary conditions, an iMPS assumes translational invariance and represents the wavefunction as a repeating tensor network that extends infinitely in both directions. This is complemented with infinite version of the TN algorithms, the iTEBD \cite{Orus_2008} and iDMRG \cite{Ostlund_1995}.
In our work we will focus on the finite version for simplicity. However, an iMPS can be used instead with minimal adaptations.

\subsection{\label{subsec:The-DMRG-algorithm}The DMRG algorithm}

\begin{figure}
    \centering
    \begin{subfigure}[b]{0.24\textwidth}
        \centering
        \includegraphics[height=0.625\linewidth]{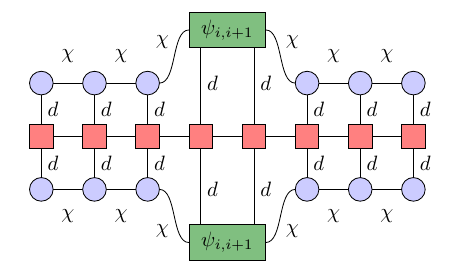}
        \caption{}
        \label{fig:2-Site-Expanded}
    \end{subfigure}\hfill
    \begin{subfigure}[b]{0.24\textwidth}
        \centering
        \includegraphics[height=0.625\linewidth]{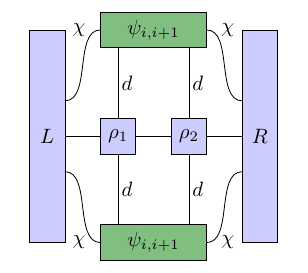}
        \caption{}
        \label{fig:2-Site-Compressed}
    \end{subfigure}
    \caption{\label{fig:2-site-operator}\justifying Two-site effective operator in the DMRG
    algorithm. The operator acts on two adjacent MPS tensors (sites $i$
    and $i+1$) during a local update step, while the surrounding environment
    tensors appearing in (a) are contracted to form the left and right effective environments in (b).
    This construction enables the variational optimization of the two-site
    block within the full many-body context. }
\end{figure}

The \emph{DMRG algorithm} variationally finds the ground state of a quantum Hamiltonian.
Within the context of the MPS ansatz, DMRG operates by sweeping back
and forth through the network of site tensors representing the state
$\ket{\Phi}$, and changing the local tensors such that they minimize
the energy $\braket{\Phi\left|H\right|\Phi}$. DMRG fixes the network
at all but one (or two) MPS tensors at a time, and searches for the
ground state of the locally-obtained Hamiltonian. This can be seen
graphically in Fig. \ref{fig:2-site-operator}.

A simple approach to find the local one (or two) site ground state,
is to construct the local operator demonstrated in Fig. \ref{fig:2-site-operator},
and to diagonalize it. Since the MPS and MPO both have bounded bond
dimensions, this local operator is of bounded size. As can be seen
in Fig. \ref{fig:2-site-operator}, the size of such operator is $\left(\chi^{2}d^{2}\times\chi^{2}d^{2}\right)$,
with $\chi$ being the state bond dimension and $d$ the physical
dimension. 

\subsection{The Lanczos Algorithm}\label{subsec:Lanczos_Algorithm}

As mentioned in Sec. \ref{subsec:The-DMRG-algorithm} above, the DMRG
algorithm requires obtaining the eigenvalues of the local Hamiltonian
in every step. Naively, one would diagonalize the $\chi^{2}d^{2}\times\chi^{2}d^{2}$
matrix. However, diagonalization is a computationally expensive operation,
with complexity of $O\left(n^{3}\right)$ for an $n\times n$ matrix in standard methods 
\cite{Golub_2013}. Therefore, diagonalization of the 2-site operator sets a
limit to the bond dimension of the MPS and MPO. This limitation can
be improved by replacing diagonalization with the \emph{Lanczos algorithm}.
The algorithm iteratively constructs a subspace of orthogonal vectors,
called a Krylov subspace, and projects the operator into this smaller space. 
This low-dimensional projection can be then diagonalized
at a lower computational cost. 
Using this diagonalization and the Krylov vectors,
we can retrieve the approximated diagolaization of the operator. 
The Lanczos algorithm is summarized in Alg. \ref{alg:Lanczos-algorithm}.

\begin{table}[tbp]
\begin{minipage}{\linewidth}
\begin{algorithm}[H]
\caption{The Lanczos algorithm}
\begin{algorithmic}[1]
    \Require{$A, \ket{q_1} \neq 0$}
    \Ensure{Eigenpair $(\theta_j, \ket{s_j})$}
    \State{$\alpha_1 \gets \braket{q_1|A|q_1}$}
    \State{$\ket{u_1} \gets A\ket{q_1} - \alpha_1\ket{q_1}$}
    \For{$j=2,...$}
        \State{Remove components from $\ket{u_j}$ to make it orthogonal to all $\left\{\ket{q_k}\right\}_{k=1}^j$, if necessary.}
    	\State{$\beta_{j} \gets \left\Vert\ket{u_{j-1}}\right\Vert$}
    	\State{$\ket{q_j} \gets \ket{u_j} / \beta_{j}$}
    	\State{$\alpha_j \gets \braket{q_j|A|q_j}$}
    	\State{$\ket{u_{j}} \gets A\ket{q_j} - \alpha_j \ket{q_j} - \beta_j \ket{q_{j-1}}$}
        \State{$T_{j,j}\gets\alpha_j$}
        \State{$T_{j-1,j}\gets\beta_j\; ; \; T_{j,j-1} \gets \beta_j^*$}
    	\State{Compute eigenpair $(\theta_j, \ket{s_j})$ of $T_j$ and check for convergence.}
    \EndFor
\end{algorithmic}
\end{algorithm}
\end{minipage}
\caption{\justifying The Lanczos algorithm constructs a Krylov subspace $\left\{ \ket{q_{j}}\right\}$
    and diagonalizes the projection of an operator $A$ into this subspace.
    While the exact solution is guaranteed only when a full space is constructed,
    typically the solution converges fast, that is, for a relatively small Krylov subspace \cite{Golub_2013}. }
\label{alg:Lanczos-algorithm}
\end{table}

The combination of the MPS framework, DMRG sweep-based optimization,
and Krylov subspace solvers such as Lanczos, results in a numerically
efficient and highly accurate approach for computing ground states
of 1D quantum systems. These techniques have since been
extended and generalized to time-evolution, calculation of excited states, and the solution ofhigher-dimensional
systems through tensor network methods.

\subsection{The GEP}\label{subsec:GEP}

Eigenvalue problems play a central role in the solution of many physical
systems, the most prominent example being the time-independent Schr\"{o}dinger equation $H\ket{\Phi}=E\ket{\Phi}$.
This problem can be formulated as searching for $\left\{ \left(\lambda_{i},\ket{\psi_{i}}\right)\right\} _{i=0}^{N}$ 
satisfying $A\ket{\psi_{i}}=\lambda_{i}\ket{\psi_{i}}$. This mathematical
problem can be generalized, by rewriting it as $A\ket{\psi_{i}}=\lambda_{i}I\ket{\psi_{i}}$, 
and substituting $I$ by some general positive definite matrix $B$, yielding the GEP,
\begin{equation}
\label{eq:GEP-definition}
A\ket{\psi_{i}}=\lambda_{i}B\ket{\psi_{i}}.
\end{equation}
The GEP arises in many physical contexts. In molecular dynamics
and lattice vibrations, vibrational modes $\ket{u}$ and their frequencies
$\omega$ are solutions to the GEP with the dynamical matrix $D$
and the mass matrix $M$: $D\ket{u}=\omega^{2}M\ket{u}$ \cite{Ashcroft_1976, Born_1996}. The GEP
also arises in photonics and plasma physics with electromagnetic wave
propagation in anisotropic media \cite{Johnson_2001, Joannopoulos_2011, Raghu_2008, Sozuer_1992, Bergmann_1982}, and in quantum chemistry
with the Roothaan equations derived from Hartree-Fock theory with
non-orthogonal atomic basis \cite{Roothaan_1951, McArdle_2020, Szabo_1996, Ford_1974}.

Naively, the GEP can be solved by transforming it to the standard eigenvalue problem:
\begin{equation}
\label{eq:GEP-naive-sol}
B^{-1}A\ket{\psi_{i}}=\lambda_{i}\ket{\psi_{i}},
\end{equation}
However, this solution depends on the invertibility of $B$, which could be problematic numerically. 
Additionally, it is computationally expensive in our context. 
Moreover, even if $A$ and $B$ are both Hermitian, $B^{-1} A$ is not.
A similar construction can be done based on the eigen-decomposition
of $B$, which is similarly computationally expensive.

We propose a protocol for solving the GEP in the MPS 
formalism, tailored for calculating the R\'{e}nyi divergence as 
presented in Sec. \ref{subsec:Renyi-divergence} below.

\subsection{R\'{e}nyi divergences and mutual information}\label{subsec:Renyi-divergence}

Quantum mutual information is a fundamental quantity in quantum information
theory that quantifies the total correlations---both classical and
quantum---between two subsystems of a larger quantum system \cite{Groisman_2005}. 
Consider a global quantum system described by a density matrix $\rho$, and
let $\rho_{AB}=\mathrm{Tr_{\overline{AB}}\left(\rho\right)}$ denote
the reduced density matrix of subsystems $A$ and $B$, obtained by
tracing out the rest of the system. 
The \emph{quantum mutual information} between $A$ and $B$ is defined as \cite{Wilde_2013}:
\begin{equation}
\label{eq:quantum-mutual-information}
I\left(A:B\right)=S\left(\rho_{A}\right)+S\left(\rho_{B}\right)-S\left(\rho_{AB}\right),
\end{equation}
where $\rho_{A}=\mathrm{Tr_{B}}\left(\rho_{AB}\right)$ and $\rho_{B}=\mathrm{Tr}_{A}\left(\rho_{AB}\right)$
are the reduced density matrices of subsystems $A$ and $B$ respectively,
and $S\left(\rho\right)$ is the von Neumann entropy \cite{Wilde_2013}:
\begin{equation}
\label{eq:von-Neumann-entropy}
S\left(\rho\right)=-\mathrm{Tr}\left(\rho\log\rho\right).
\end{equation}
The von Neumann entropy quantifies the amount of uncertainty, or ``mixedness'',
of a quantum state and generalizes the classical Shannon entropy to
the quantum setting. Quantum mutual information is non-negative and
vanishes if and only if the state is a product state, i.e., $\rho_{AB}=\rho_{A}\otimes\rho_{B}$,
indicating no correlation between $A$ and $B$. Beyond the von Neumann
entropy, a broader, classical or quantum, class of entropic measures is given by R\'{e}nyi entropies,
which introduce a parameter $\alpha>0$ and $\alpha\ne1$. The \emph{R\'{e}nyi
entropy of order $\alpha$} for a density matrix $\rho$ is defined
as \cite{Wilde_2013}:
\begin{equation}
\label{eq:Renyi-entropy}
S_{\alpha}\left(\rho\right)=\frac{1}{1-\alpha}\log\mathrm{Tr}\left(\rho^{\alpha}\right),
\end{equation}
which reduces to the von Neumann entropy in the limit $\alpha\to1$.
Analogously, the \emph{R\'{e}nyi mutual information of order $\alpha$}
can be defined as \cite{Wilde_2013}:
\begin{equation}
\label{eq:renyi-mutual-information}
I_{\alpha}\left(A:B\right)=S_{\alpha}\left(\rho_{A}\right)+S_{\alpha}\left(\rho_{B}\right)-S_{\alpha}\left(\rho_{AB}\right).
\end{equation}
For integer $\alpha$, these measures require the evaluation of powers and traces of the density matrices, making them calculable.
$S_\alpha$ and $I_\alpha$ have been calculated in several physical settings, including conformal field theory 
\cite{Pasquale_2004, Calabrese_2009}, free fermions \cite{Bernigau_2015}, MPDO states \cite{Pirvu_2010},
and using quantum Monte Carlo methods \cite{Cirac_2010, Hastings_2010, Humeniuk_2012, Grover_2013}.
These measures have been shown to characterize important physical phenomena \cite{Alcaraz_2014, Jean_Marie_2014, Singh_2011, Banuls_2017}. 
However, in general these measures do not have an operational meaning and may even become negative [18], making them inadequate as a mutual information measures.

Another measure of mutual information can be considered, arising from an equivalent
definition of the von Neumann quantum mutual information \cite{Khatri_2024, Tomamichel_2016}:
\begin{equation}
\label{eq:relative-entropy}
I\left(A:B\right)=D\left(\rho_{AB}\Vert\rho_{A}\otimes\rho_{B}\right),
\end{equation}
with $D\left(\rho\Vert\sigma\right)$ being the Umegaki relative entropy \cite{Wilde_2013}:
$D\left(\rho\Vert\sigma\right)=\mathrm{Tr}\left(\rho\log\rho-\rho\log\sigma\right)$.
This definition of the quantum mutual information can be generalized using R\'{e}nyi relative entropies, 
yielding the \emph{R\'{e}nyi divergence of order $\alpha$} \cite{Khatri_2024, Tomamichel_2016}. 
These measures are non-negative, and cannot increase under local operations. 
Following Ref. \cite{Scalet_2021}, we will focus on the $\alpha \to \infty$ limit, the so-called \emph{maximal R\'{e}nyi divergence}, given by \cite{Datta_2009}:
\begin{align}
\label{eq:maximal-divergence}
D_{\infty}\left(\rho\Vert\sigma\right) & =\log\inf\left\{ \lambda:\rho\leq\lambda\sigma\right\} \nonumber \\
 & =\log\inf\left\{ \lambda:\inf_{\ket{\psi}}\braket{\psi\left|\lambda\sigma-\rho\right|\psi}\geq0\right\},
\end{align}
which yields the maximal R\'{e}nyi mutual information:
\begin{equation}
\label{eq:maximal-Renyi-information}
I_{\infty}\left(A:B\right)=D_{\infty}\left(\rho_{AB}\Vert\rho_{A}\otimes\rho_{B}\right).
\end{equation}
In our work, we will focus on the efficient calculation of this measure,
by the solution of a GEP as described below.

In order to calculate the maximal divergence efficiently, we start from the second line of Eq. \eqref{eq:maximal-divergence}.
For finite dimensional $\rho$, $\sigma$, the value 0 is achievable, and we are left with the
generalized eigenvalue problem: $\rho \ket{\psi} = \lambda \sigma \ket{\psi}$. 
This can be written as \cite{Ghojogh_2023}:
\begin{equation}
\label{eq:lmb}
\lambda_{\infty}=\max_{\ket{\tilde{\psi}}}\braket{\tilde{\psi}|\sigma^{-1/2}\rho\sigma^{-1/2}|\tilde{\psi}},
\end{equation}
by transforming the generalized eigenvector: $\ket{\tilde{\psi}}=\sigma^{1/2}\ket{\psi}$.

Let us note that since $\sigma$ is a positive semidefinite matrix and could be singular, it is not trivial
that $\sigma^{-1}$ can be calculated.
Additionally, the effects of zero (or near zero) eigenvalues, which would cause this measure to
diverge, must be explicitly regularized, to not significantly affect the calculation. 
This is shown in appendix \ref{sec:Irrelevance-of-Kernel}. 

A note regarding notations: In our work we will use $\ket{\Phi}$ to denote the state of the entire physical system, which is a pure state represented by the full MPS. 
We will use $\ket{\psi}$ to denote vectors in the space of $\rho$ and $\sigma$. 
These should not be confused, as they serve different physical meaning and are taken from different Hilbert spaces.

\section{Calculating the maximal divergence for Subsystems at the Edges}\label{sec:System-Edge-Calculation}

\begin{figure}
    \includegraphics[width=\linewidth]{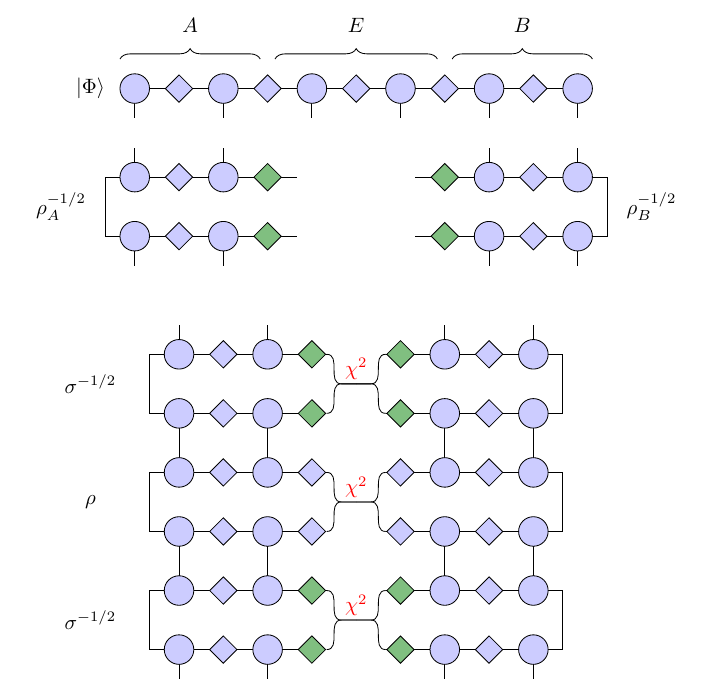}
    \caption{\label{fig:edge_subsystems}\justifying Construction of the subsystem density
    matrix can be done by using the state MPS structure. Site matrices
    (circles) are saved in left and right canonical form. Schmidt values
    (diamonds) are saved as well. Using the Schmidt decomposition: $\ket{\Phi}=\sum_{i}S_{i}\ket{i_{A}}\bra{i_{E,B}}$
    the reduced density matrix $\rho_{A}$ is simply $\rho_{A}=\sum_{i}S_{i}^2\ket{i_{A}}\bra{i_{A}}$.
    Therefore, the eigenvector decomposition of $\rho_{A}$ is given
    by the left canonical matrices of A, with eigenvalues being the Schmidt
    values at the subsystem edge (green diamonds). 
    In terms of Eq. \eqref{eq:MPS-definition-with-S}, the eigenvalues of $\rho_A$ are the 
    diagonal elements of the matrix $\Lambda^i$
    at its edge, with the corresponding eigenvectors $\sum_{s_i} \Gamma^{s_1}\Lambda^1\cdots\Gamma^i$.
    Consequently, manipulations of $\rho_{A}$ such as calculation of $\rho_{A}^{-1/2}$ can be done
    by raising the Schmidt values from the diagonal of $\Lambda^i$ to power $-1/2$; when contracting the bra and ket MPSs, we then square these values. (green diamonds),
    giving $\rho_{A}^{-1/2},\rho_{B}^{-1/2}$ (top). Chaining these gives
    $\sigma^{-1/2}=\rho_{A}^{-1/2}\otimes\rho_{B}^{-1/2}$, which can be
    used to construct $\sigma^{-1/2}\rho\sigma^{-1/2}$ (bottom).}
\end{figure}

We begin by referring to the case in which the subsystems A,B are each contiguous and lie
on the edges of the system, as depicted in Fig. \ref{fig:edge_subsystems} (top).
Such a case allows us to utilize the MPS orthogonality structure and
manipulate the reduced density matrices. Using this method we are
able to calculate non trivial operators, such as operators inverse. 

The formulation of the problem as in Eq. $\eqref{eq:lmb}$ has the
benefit that it can be calculated using a single DMRG execution. 
To calculate $\sigma^{-1/2}$, we need to access $\sigma$'s eigenvector
decomposition and to manipulate its eigenvalues. 
Thanks to the MPO structure of $\sigma$ derived from the system MPS, constructing
$\sigma^{-1/2}$ can be done efficiently by accessing the orthogonality
center values at the ``inner'' edges of subsystems $A$, $B$. 
By Eq. (4), the $\Lambda^i$ at the edge of subsystem A/B contain the corresponding singular (Schmidt) values,
whose squares are the eigenvalues of the corresponding reduced density matrix $\rho_{A/B}$, 
with the eigenvectors being the corresponding contracted canonical tensors.
To calculate $\rho_{A/B}^{-1/2}$, the inverse of the square roots of the Schmidt values 
are calculated, while leaving the isometric MPS matrices the same; when contracting the bra and ket MPSs, the eigenvalues of $\rho_{A/B}^{-1/2}$ thus become the inverses of the Schmidt values.
As mentioned before, the Schmidt values are regularized to prevent (near-) zero eigenvalue to affect the calculation.
This regularization is detailed in Sec. \ref{subsec:Technical-Implementation-Details}.
This calculation yields the inverse to the subsystems density matrix $\rho_{A}^{-1/2},\rho_{B}^{-1/2}$, from which $\sigma^{-1/2}=\rho_{A}^{-1/2}\otimes\rho_{B}^{-1/2}$ readily follows.
Eq. (\ref{eq:lmb}) may be then computed by applying the DMRG algorithm
to the matrix as demonstrated in Fig. \ref{fig:edge_subsystems} (bottom).

This calculation relies on the structure of edge subsystems. The calculation
of $\sigma^{-1/2}$ is possible thanks to the immediate accessibility
of $\sigma$'s eigenvector decomposition, allowing us to manipulate
the eigenvalues directly. Since this is not the case generally, the
general case will require a different calculation method. Additionally,
this operator is of large size, with bond dimension $\chi^{6}$ compared
to the state bond dimension $\chi$, as demonstrated in Fig. \ref{fig:edge_subsystems}
(bottom). While these issues limit the usability of this method, this
construction is simple to implement and therefore is important as
a benchmark for the more general approach presented below.

\section{Calculating the maximal divergence in the general case}\label{sec:general-case-calculation}

We proceed from the example of edge subsystems to the general case. In
this context, calculations of operators such as $\sigma^{-1/2}$ is
no longer efficiently feasible.
We propose to calculate the maximal divergence by adapting the DMRG
algorithm, and change its objective:

\begin{equation}
\label{eq:adapted_dmrg}
\lambda=\min_{\ket{\psi}}\frac{\braket{\psi|\rho|\psi}}{\braket{\psi|\psi}}\to\lambda_{\infty}=\min_{\ket{\psi}}\frac{\braket{\psi|\rho|\psi}}{\braket{\psi|\sigma|\psi}}.
\end{equation}
This may be achieved by adapting the standard Lanczos algorithm, used
within the core DMRG iteration for solving the two-site eigenvalue
problem, to its generalized form suitable for the generalized eigenvalue
problem (GEP).

In Sec. \ref{subsec:Solving-the-GEP} we introduce a high-level view
of the generalized Lanczos algorithm, in Sec. \ref{subsec:Technical-Implementation-Details}
we elaborate on the technical details and the required adjustments
to the generalized Lanczos algorithm for its successful numerical
calculation, and in Sec. \ref{subsec:Application-to-DMRG} we present
the adaptation to the DMRG algorithm, and discuss its difficulties.

\subsection{Solving the Generalized Eigenvalue Problem}\label{subsec:Solving-the-GEP}

The adapted Lanczos algorithm to the generalized eigenvalue problem may be found in Alg. \ref{alg:Generalized-Lanczos}. 
This algorithm is our adaptation of the algorithm appearing in Ref. \citep{Parlett_1980} 
to improve its stability, as discussed in detail in appendix \ref{sec:Generalized-Lanczos-eq}.

\begin{table}[tpb]
\begin{minipage}{\linewidth}
\begin{algorithm}[H]
\caption{Generalized Lanczos}
\begin{algorithmic}[1]
    \Require{$A, M, \ket{u_1} \neq 0$}
    \Ensure{Generalized eigenpair $(\theta_j, \ket{s_j})$}
    \For{$j=1,2,...$}
        \State{Remove components from $\ket{u_j}$ to make it M-orthogonal to all $\left\{\ket{q_k}\right\}_{k=1}^j$, if necessary.}
    	\State{$\ket{q_j} \gets \ket{u_j} / \sqrt{\braket{u_j|M|u_j}}$}
    	\State{$\alpha_j \gets \braket{q_j|A|q_j}$}\label{alg:line:gen-Lanczos-alpha}
    	\State{$\beta_{j} \gets \braket{q_{j-1}|A|q_j}$}\label{alg:line:gen-Lanczos-beta}
    	\State{$\ket{r_{j+1}} \gets A\ket{q_j} - \alpha_j M \ket{q_j} - \beta_j M \ket{q_{j-1}}$}\label{alg:line:gen-Lanczos-r}
    	\State{Solve $M\ket{u_{j+1}}=\ket{r_{j+1}}$ for $\ket{u_{j+1}}(=\beta_{j+1}\ket{q_{j+1}})$}
    	\State{Compute eigenpair $(\theta_j, \ket{s_j})$ of $T_j$ and check for convergence.}
    \EndFor
\end{algorithmic}
\end{algorithm}
\end{minipage}
\caption{\justifying The generalized Lanczos algorithm, adjusted from Alg. \ref{alg:Lanczos-algorithm} for GEP optimization. 
    This formulation of the algorithm is a modification of the version given by Ref. \cite{Parlett_1980}, as discussed in appendix \ref{sec:Generalized-Lanczos-eq}.
    M-orthogonality, used in this algorithm, is defined by: $\left\vert\braket{q_i\left\vert M \right\vert q_j}\right\vert=\delta_{ij}$.}
\label{alg:Generalized-Lanczos}
\end{table}

In this formulation, Krylov space vectors are $\sigma$-orthonormal:
$\braket{q_{i}|\sigma|q_{j}}=\delta_{ij}$. Additionally,
the generalized algorithm differs from the standard in that the
calculation of the Krylov space vectors now requires solving a linear
equation, $\sigma\ket{a}=\ket{b}$. 
However, relying on the MPO structure, one is able to access only positive integer powers of $\sigma$. 
To solve these linear equation using such powers of $\sigma$,
we employed the Conjugate Gradient (CG) algorithm, presented in Alg. \ref{alg:Conjugate-Gradient} \cite{Hestenes_1952}. 

\begin{table}[t!]
\begin{minipage}{\linewidth}
\begin{algorithm}[H]
\caption{Conjugate Gradient}
\begin{algorithmic}[1]
    \Require{$A, \ket{b}, \ket{x_0}$}
    \Ensure{$\ket{x}$ such that $A\ket{x}-b\approx 0$}
    \State{$\ket{r_0}\gets\ket{b}-A\ket{x_0}$}
    \State{if $\ket{r_0}$ is sufficiently small, return $\ket{x_0}$}
    \State{$\ket{p_0}\gets\ket{r_0}$}
    \For{$j=0,1,...$}
    	\State{$\alpha_j\gets\frac{\braket{r_j|r_j}}{\braket{p_j|A|p_j}}$}
    	\State{$\ket{x_{j+1}}\gets\ket{x_j}+\alpha_j\ket{p_j}$}
    	\State{$\ket{r_{j+1}}\gets\ket{r_j}-\alpha_jA\ket{p_j}$}
    	\State{if $\ket{r_{j+1}}$ is sufficiently small, exit loop}
    	\State{$\beta_j\gets\frac{\braket{r_{j+1}|r_{j+1}}}{\braket{r_j|r_j}}$}
    	\State{$\ket{p_{j+1}}\gets\ket{r_{j+1}}+\beta_j\ket{p_j}$}
    \EndFor
    \State{return $\ket{x_{j+1}}$}
\end{algorithmic}
\end{algorithm}
\end{minipage}
\caption{\justifying The Conjugate Gradient algorithm \cite{Hestenes_1952} is an iterative algorithm for the solution
    of a system of linear equations. The algorithm constructs a Krylov subspace of the vectors $\ket{r_j}$ which are
    orthonormal, and a Krylov subspace of the vectors $\ket{p_j}$ which are orthonormal with respect to the inner product $\braket{p_j|A|p_j}$. Both sets span the same subspace. At each iteration, the approximated solution $\ket{x_j}$
    is the projection of the theoretical solution to this Krylov subspace.}
\label{alg:Conjugate-Gradient}
\end{table}

\subsection{Technical Implementation Details}\label{subsec:Technical-Implementation-Details}

In this section, we dive into some of the technical aspects of the
implementation of the generalized Lanczos algorithm. We cover the
nontrivial details necessary for implementation of the algorithm
within the context of our objective, which may be applicable in other
physical systems.

The generalized Lanczos and the CG algorithms both require positive
definite matrices, rather than positive semi definite matrices.
Since tensor networks are an approximation reliant on the idea of
reducing the rank of matrices by keeping only the largest Schmidt values,
by construction $\sigma$ and $\rho$ have a large kernel (null space). 
As discussed in Appendix A, it is necessary to introduce a regulator that enforces positivity,
specifically for $\sigma$ which is effectively inverted within the
algorithms. This was done by using $\tilde{\sigma}=\sigma+\varepsilon I$.
To solve for the linear equation within the algorithm, the CG algorithm
is used with each application of $\sigma$ replaced with $\tilde{\sigma}$,
effectively calculating $\tilde{\sigma}^{-1}$. While more subtle
regulators can be introduced to provide other forms of $\tilde{\sigma},\tilde{\sigma}^{-1}$,
to maintain the $\sigma$-orthogonality of the Krylov subspace and
tri-diagonality of the projected $\rho$, the relation $\tilde{\sigma} \tilde{\sigma}^{-1} = I$ must be maintained. 
This rules out regulators such as $\tilde{\sigma}^{-1} = \sigma/\left(\sigma^2 + \epsilon^2\right)$, which has no inverse, 
and thus no well-behaved corresponding $\tilde{\sigma}$.
Importantly, the regularization introduces an error into
the calculation, and so $\varepsilon$ is chosen to be the required
overall accuracy.

\subsection{\label{subsec:Application-to-DMRG}The generalized DMRG algorithm}

Using this modified Lanczos algorithm, we can turn to adapting the
DMRG algorithm to the generalized eigenvalue problem. 
This is straightforward in principle: The algorithm maintains 2 MPOs $\left(\rho,\sigma\right)$
and at each iteration 2-site operators are constructed for both
MPOs, and passed on to the generalized Lanczos algorithm. 
This results in a greedy algorithm that maximizes the ratio in Eq. (\ref{eq:adapted_dmrg}) site by site, as
described in Alg. \ref{alg:Generalized-DMRG}.

\begin{table}[t!]
\begin{minipage}{\linewidth}
\begin{algorithm}[H]
\caption{Generalized DMRG}
\begin{algorithmic}[1]
    \Require{$\rho, \sigma, \{\psi_i\}$}
    \Ensure{Generalized eigenpair $(\lambda, \ket{\psi^\prime})$}
    \State{Calculate initial $\lambda=\braket{\psi|\rho|\psi}/\braket{\psi|\sigma|\psi}$}
    \For{$s = 1,2,...$}
    	\For{$i = 1,2,...,N-2,N-1,N-2,...,2,1$}
    		\State{$\rho_2, \sigma_2 \gets$ 2-site operators of sites $i,i+1$ for $\rho,\sigma$}
    		\State{$(\lambda^\prime,\psi_{i,i+1}^\prime)\gets\mathrm{GeneralizedLanczos}(\rho_2,\sigma_2,\psi_{i,i+1})$ }
    		\State{$\psi_{i,i+1}\gets\psi_{i,i+1}^\prime$}
    		\State{$\lambda\gets\lambda^\prime$}
    	\EndFor
    	\State{Check for convergence of $\lambda$, if converged return $(\lambda, \{\psi_i\})$}
    \EndFor
\end{algorithmic}
\end{algorithm}
\end{minipage}
\caption{\justifying Our generalized DMRG algorithm}
\label{alg:Generalized-DMRG}
\end{table}

If the system possesses conserved charges, they can be used in the context of MPSs to reduce 
computational complexity.
For example, one may simplify the calculation of $H\ket{\Phi}$ when $H$ is block diagonal and 
$\ket{\Phi}$ has a well defined charge. 
In the case of the DMRG algorithm, since
numerical diagonalization is replaced with the Lanczos algorithm which
uses moments of the operator, the resulting ground state is limited
to the subspace of the charge to of the original state. For instance,
given $H$ which describes a spin system in a ferromagnetic regime,
if the initial vector of the DMRG is that of an anti-ferromagnet, the resulting
ground state is still limited to the zero total spin subspace. 
Similar observations apply to our generalized DMRG.

A noteworthy complication that arises in the context of the GEP as it is applied to the DMRG algorithm for
the calculation of the maximal divergence, is that the spin
of the generalized eigenvector does not have to match with that of
the system state. For example, an anti-ferromagnetic system, with
zero total spin, could have a solution to Eq. (\ref{eq:maximal-divergence})
with $\ket{\psi}$ having nonzero spin, e.g., when the subsystems each contain an odd number of sites.

For the implementation of the algorithms we used the Python TN framework TeNPy \cite{tenpy_1, tenpy_2}.
This framework was used for its efficiency and ease of use in the manipulations of the tensors.
While TeNPy offers implementations for many algorithms, we chose to implement some algorithms which exist within TeNPy on our own.
This was done to allow us to focus solely on the necessary features and to give us more control over the design.
These implementations can be easily adapted to be incorporated into the TeNPy framework. 

\subsection{\label{subsec:complexity}The complexity of the algorithm}
The convergence rate of algorithm \ref{alg:Generalized-Lanczos} is discussed in appendix \ref{sec:gen-lanczos-correctness}. 
In practice, however, the bounds provided there are often unhelpful as we do not have access to the eigenvalue decomposition of the matrices.
Instead, we can view the complexity of Alg. \ref{alg:Generalized-Lanczos}, \ref{alg:Conjugate-Gradient} and \ref{alg:Generalized-DMRG} 
independently, and these can then be compared with the standard DMRG.
The given parameters of the problem are the bond dimension of the MPOs $\chi_1$, the target bond dimension of the MPS which 
solves the GEP $\chi_2$, the total number of sites $N$, and the physical dimension $d$. 

In the case of calculating the maximal R\'enyi divergence, the total number of sites $N$ is the number of sites in both subsystems,
not the overall size of the physical system.
Additionally, the MPOs are density matrices and therefore their bond dimension $\chi_1$ is derived from the bond dimension of
the system state $\chi_s$ by: $\chi_1=\chi_s^2$.
The memory complexity in this case is dominated by the representation of the density matrices, which can become prohibitively large.
Each site tensor of the density matrices is of size $d^2\chi_{1}^2$, and the overall memory complexity is $Nd^2\chi_{1}^2$.
In terms of the bond dimension of the system state $\chi_s$, the memory complexity is $Nd^2\chi_s^4$.
The computational complexity of the generalized DMRG differs from the standard algorithm only in the invocation of the Conjugate Gradient 
algorithm, appearing in Alg. \ref{alg:Conjugate-Gradient}. The CG algorithm computes matrix-vector product similarly to the generalized 
and standard Lanczos, multiplying the overall complexity by the average number of CG iterations per execution, which in our calculations was approximately 20.

\section{Results from physical models}\label{sec:results}

\begin{figure}[ht]
\includegraphics[width=\linewidth]{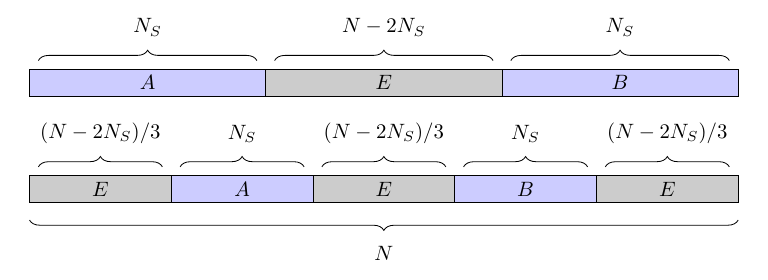}
\caption{\label{fig:arch}\justifying Two types of subsystems architectures were chosen to be studied.
Top: An `AEB' architecture where the subsystems are equally sized and are at the edges of the total system. 
This structure is discussed in Sec. \ref{sec:System-Edge-Calculation}, allowing an efficient calculation of 
the maximal divergence using a special case method.
Bottom: An `EAEBE' architecture where the subsystems are equal in size and lie at equal distances from each other and from the edges.}
\end{figure}

\begin{figure}[ht]
\includegraphics[width=\linewidth]{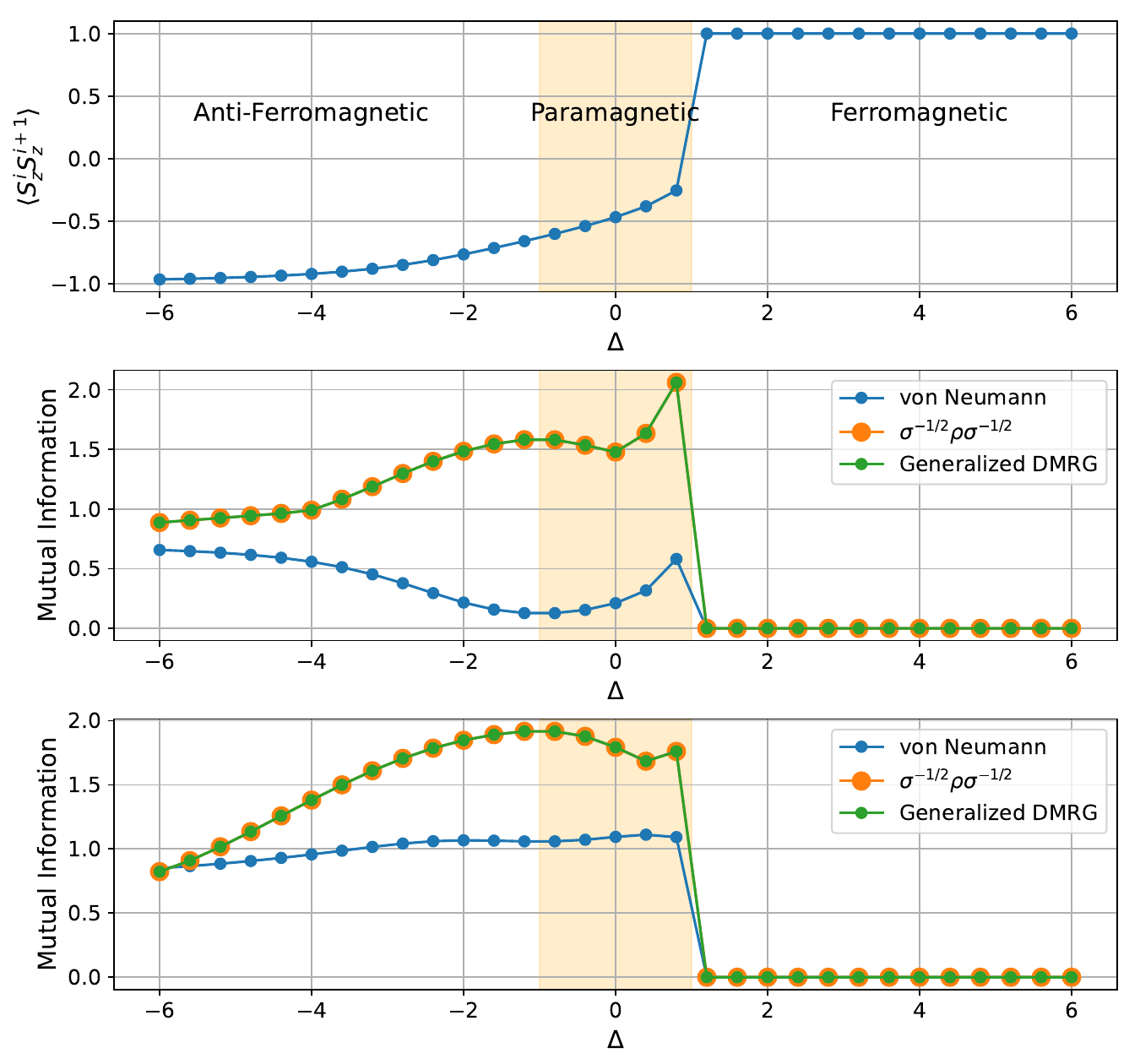}
\caption{\label{fig:XXZ-information-comparison}\justifying Testing our method on the XXZ system
for fixed $J=1,h=0$, for a system of $N=10$ sites, and subsystems with $N_S=3$ sites.
Top: Calculation of $\braket{S_{i}^{z}S_{i+1}^{z}}$ in the ground state, 
displaying the XXZ phases: At low $\Delta$ the correlation is large
and negative, meaning an anti-ferromagnet. At high $\Delta$ the correlation
is high and positive, meaning a ferromagnet. In between there is a region of a paramagnetic phase.
Center: Calculation in the `AEB' subsystems architecture, shown in Fig. \ref{fig:arch}, displaying the von Neumann mutual 
information (blue), the maximal R\'{e}nyi divergence (orange, green) calculated using the 
special case method discussed in Sec. \ref{sec:System-Edge-Calculation} (orange) and 
the general case method discussed in Sec. \ref{sec:general-case-calculation} (green). 
The mutual information and the R\'{e}nyi divergence are not continuous at the paramagnetic-ferromagnetic phase transition
and share asymptotic behavior, as explained in the main text, but otherwise differ.
Bottom: Calculation in the `EAEBE' subsystems architecture, displaying the 
von Neumann mutual information and the maximal R\'{e}nyi divergence (blue, orange) 
calculated using brute force diagonalization of the density matrices, 
and the maximal R\'{e}nyi divergence calculated using the general case method discussed in 
Sec. \ref{sec:general-case-calculation} (green). 
The maximal divergence shows a similar correlation to the mutual information as in the case of the `AEB` architecture.
}
\end{figure}

\begin{figure*}[ht]
    \centering
    \begin{subfigure}[b]{0.33\textwidth}
        \centering
        \includegraphics[width=\textwidth]{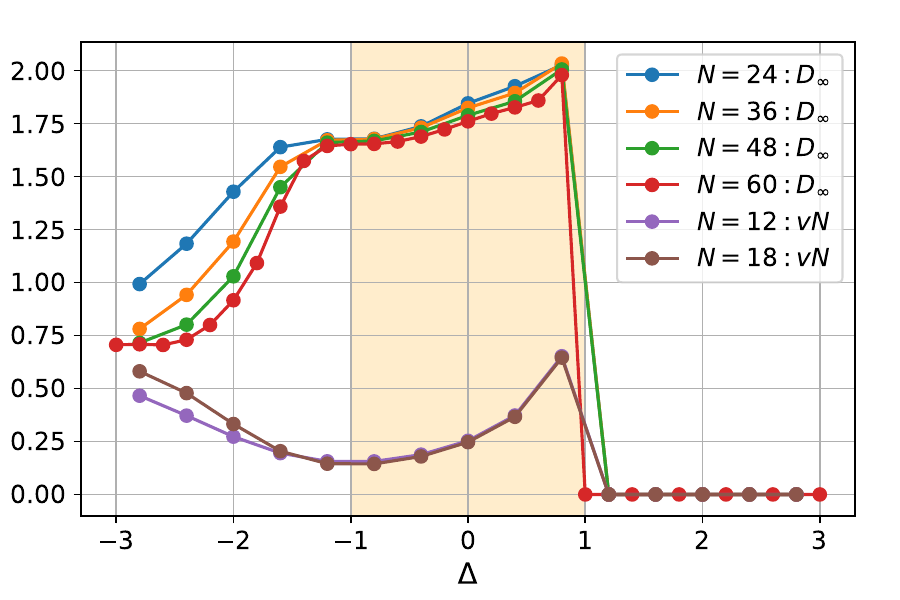}
        \caption{}
        \label{fig:AEB-phases}
    \end{subfigure}\hfill
        \begin{subfigure}[b]{0.33\textwidth}
        \centering
        \includegraphics[width=\textwidth]{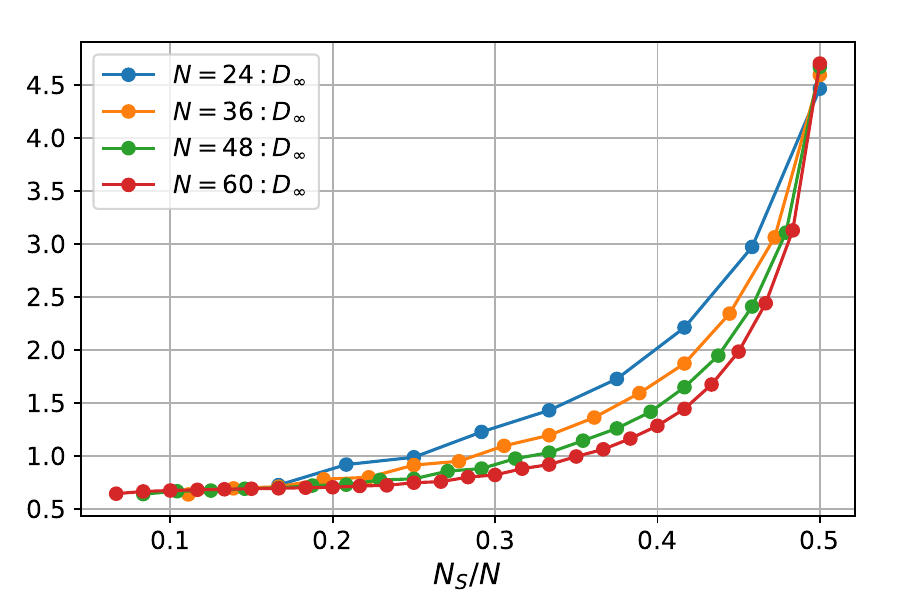}
        \caption{}
        \label{fig:AEB-delta-2}
    \end{subfigure}\hfill
    \begin{subfigure}[b]{0.33\textwidth}
        \centering
        \includegraphics[width=\textwidth]{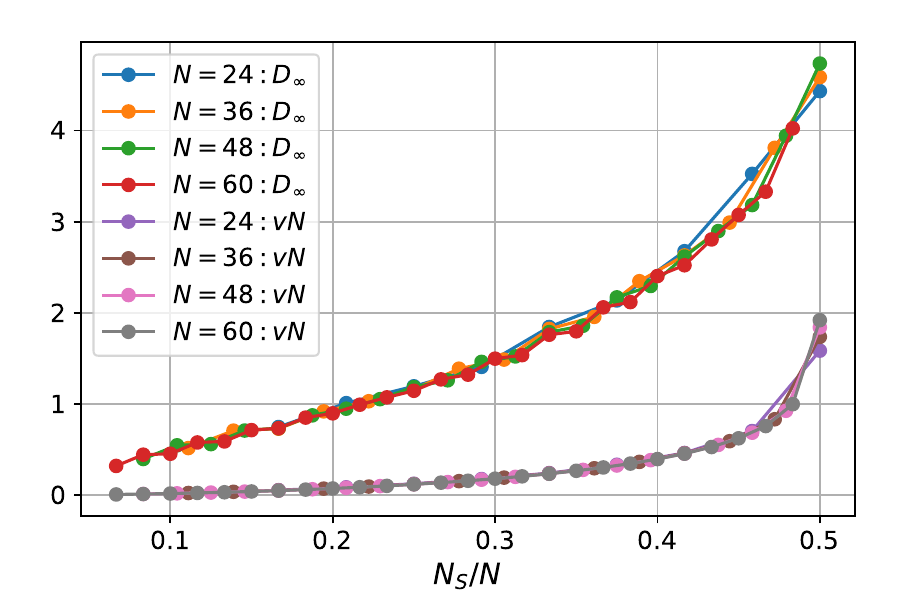}
        \caption{}
        \label{fig:AEB-delta-0}
    \end{subfigure}

    \vspace{1em}
    \begin{subfigure}[b]{0.33\textwidth}
        \centering
        \includegraphics[width=\textwidth]{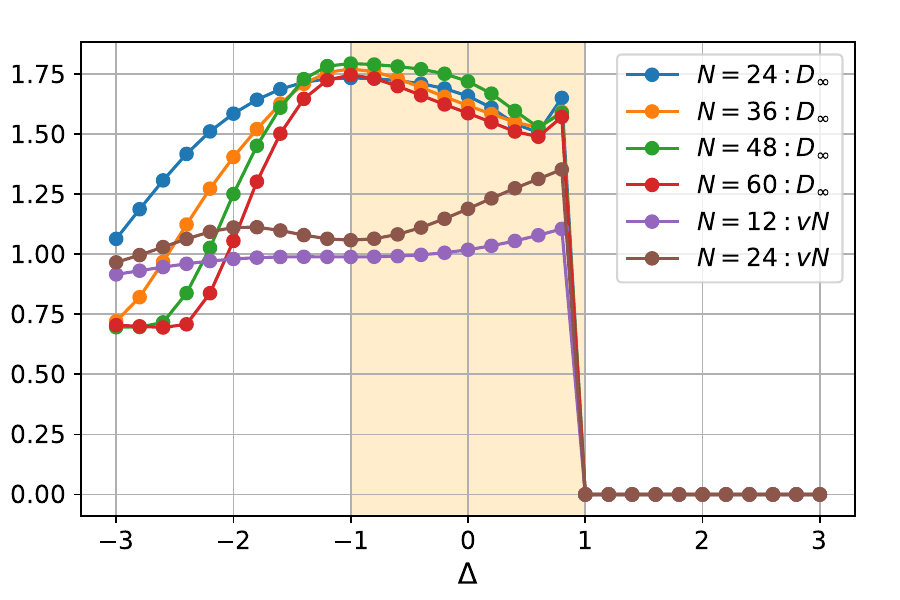}
        \caption{}
        \label{fig:EAEBE-phases}
    \end{subfigure}\hfill
    \begin{subfigure}[b]{0.33\textwidth}
        \centering
        \includegraphics[width=\textwidth]{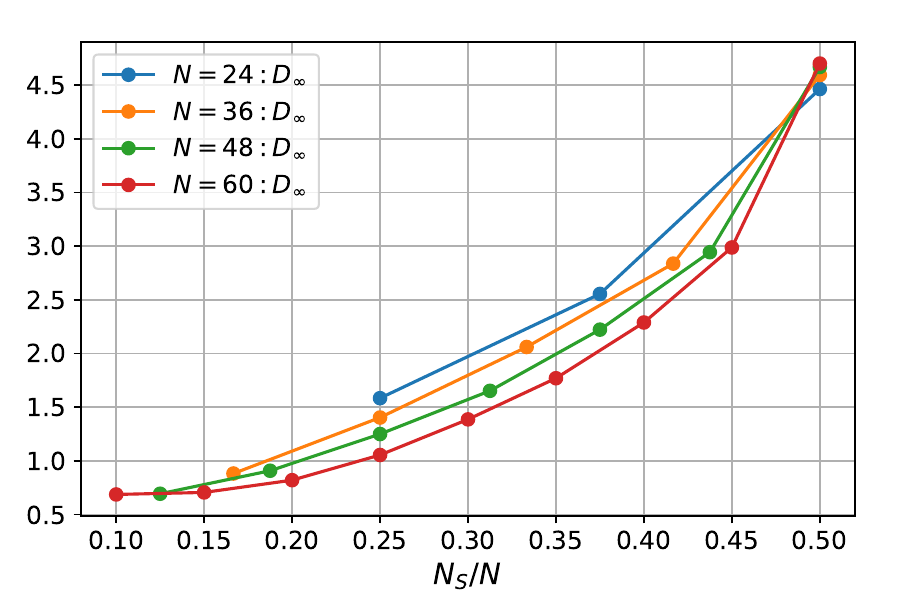}
        \caption{}
        \label{fig:EAEBE-delta-2}
    \end{subfigure}\hfill
    \begin{subfigure}[b]{0.33\textwidth}
        \centering
        \includegraphics[width=\textwidth]{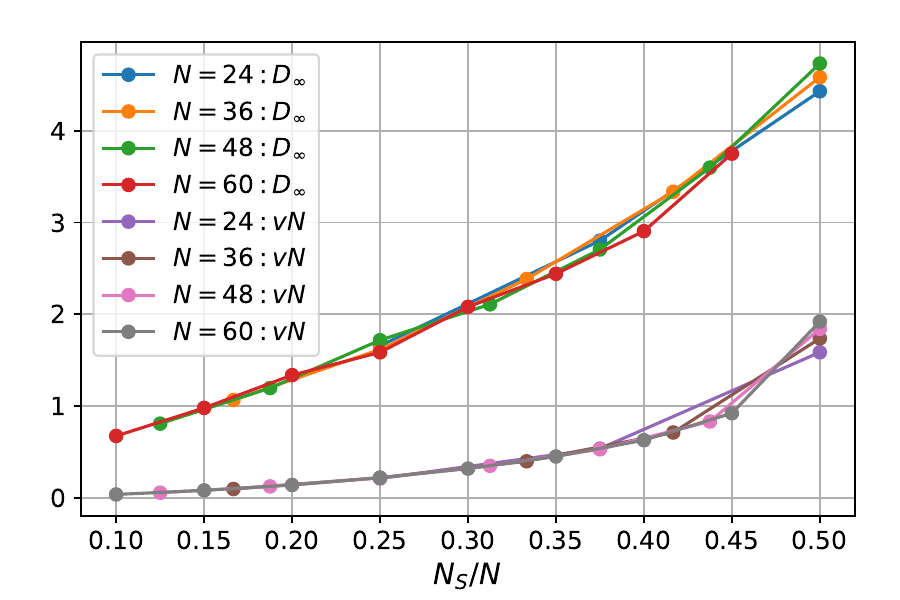}
        \caption{}
        \label{fig:EAEBE-delta-0}
    \end{subfigure}
    \caption{\justifying Results of calculating the maximal R\'enyi divergence in the XXZ chain model.
    Top row: Subsystems are equal in size and at the opposite edges of the system (Fig. \ref{fig:arch}, top panel). 
    Bottom row: Subsystems are equal in size and equally spaced from each other and from the system edges (Fig. \ref{fig:arch}, bottom panel).
    (a, d) Dependence of the R\'{e}nyi maximal divergence and von Neumann mutual information on $\Delta$ across the phase diagram of the system (depicted as in Fig. \ref{fig:XXZ-information-comparison}). The size of each subsystems is a third of the total system size.
    (b, e) The R\'{e}nyi maximal divergence as function of the subsystems sizes at $\Delta=-2$ and different system sizes.
    (c, f) The R\'{e}nyi maximal divergence as function of the subsystems sizes at $\Delta=0$ and different system sizes. The maximal divergence is plotted along the von Neumann mutual information, which is calculated using a free fermions solution.}
    \label{fig:All-Results}
\end{figure*}

To test these methods for calculating the maximal R\'{e}nyi divergence,
as well as to examine the behavior of this quantity in a physical system, we
chose the paradigmatic XXZ chain model \cite{Franchini_2017}:

\begin{equation}
\label{eq:XXZ}
H=-J\sum_{i=1}^{N}\left(S_{i}^{x}S_{i+1}^{x}+S_{i}^{y}S_{i+1}^{y}+\Delta S_{i}^{z}S_{i+1}^{z}\right)-2h\sum_{i=1}^{N}S_{i}^{z}.
\end{equation}
The parameters of the XXZ chain model are the exchange coupling $J$, the anisotropy parameter $\Delta$
and an external longitudinal magnetic field $h$.

In our calculations, we focus on the case of $h=0$ and vary $\Delta$, as well as the geometry of the subsystems.
Setting w.l.o.g $J>0$, the model exhibits three phases: 
A gapless paramagnetic phase at $|\Delta|<1$, a gapped antiferromagnetic phase at $\Delta<1$ and a gapped 
ferromagnetic phase at $\Delta>1$ \cite{Franchini_2017}.
In the following calculations we used bond dimensions up to $\chi_s=\chi_2=64$, $\chi_1=\chi_s^2$.

Choosing the subsystems to be at the edges of the system, as described in Sec. 
\ref{sec:System-Edge-Calculation} (Fig. \ref{fig:arch}, top panel), we can 
compare both methods of calculation of the maximal R\'{e}nyi divergence, described in Sec. 
\ref{sec:System-Edge-Calculation} and \ref{sec:general-case-calculation}, 
as well as compare their results to the von Neumann mutual information.
We start by also choosing a small system size, so that $\sigma^{-1/2}\rho\sigma^{-1/2}$ could be calculated
and diagonalized by brute force for arbitrary subsystems, and the results could be used to verify those of
the more efficient algorithms we developed.
As can be seen in Fig. \ref{fig:XXZ-information-comparison}, the results for the maximal divergence agree between 
the diagonalization of $\sigma^{-1/2}\rho\sigma^{-1/2}$ as described in Sec. \ref{sec:System-Edge-Calculation} (orange) and the 
generalized DMRG as described in Sec. \ref{sec:general-case-calculation} (green). 
This can be seen both in the `AEB' (Fig. \ref{fig:XXZ-information-comparison}, center panel) where the construction
of $\sigma^{-1/2}\rho\sigma^{-1/2}$ is efficient, and in the `EAEBE' (Fig. \ref{fig:XXZ-information-comparison}, bottom panel) 
architectures, where the calculation is inefficient but calculable at a small system size.

Our algorithms can be further tested in different settings. 
At larger subsystem sizes the von Neumann mutual information is no longer calculable, and if choosing to
study subsystems other than at the edges of the system, the method discussed in Sec. \ref{sec:System-Edge-Calculation} 
is no longer applicable.
In the special case of $\Delta=0$, the model reduces to that of free fermions.
As such, von Neumann entropies can be efficiently calculated by calculating the single-particle
correlations of the known solutions \cite{Peschel_2003, klich_2002}.
Using the entropies, the von Neumann mutual information can be calculated using Eq. \eqref{eq:quantum-mutual-information}. 
However, The maximal divergence cannot be directly calculated using such methods.

Comparing the maximal divergence with the von Neumann mutual information in Fig. 
\ref{fig:XXZ-information-comparison}, \ref{fig:AEB-phases} and \ref{fig:EAEBE-phases}, 
the results show that the maximal divergence reaches a maximal value along with the von Neumann mutual information, 
and then drops abruptly at the paramagnetic-ferromagnetic phase transition.
The maximal divergence asymptotically converges with the mutual information as $|\Delta|$ is increased, as the ground state 
becomes closer to the standard anti-ferromagnetic and ferromagnetic states for finite systems, 
$\frac{1}{\sqrt{2}}\left(\ket{\uparrow\downarrow\cdots} + \ket{\downarrow\uparrow\cdots}\right)$ and
$\ket{\uparrow\uparrow\cdots}$ (or $\ket{\downarrow \downarrow \cdots}$), respectively.
In these cases, both measures give the log of the number of possible states, $0$ in the 
ferromagnetic case and $\mathrm{ln}\left(2\right)$ in the anti-ferromagnetic case.
While the measures share the asymptotics and display similar behavior at the paramagnetic to ferromagnetic transition, 
as just stated, at the transition between the anti-ferromagnetic ($\Delta<-1$) 
to the paramagnetic phase ($-1\leq\Delta\leq1$) the maximal divergence displays an opposite behavior 
relative to the mutual information: The former exhibits a local maximum, while the latter a local minimum.
This shows a qualitative difference between the maximal divergence and the mutual information.

In Fig. \ref{fig:AEB-delta-0} and \ref{fig:EAEBE-delta-0} the maximal divergence is plotted along the
von Neumann mutual information, calculated using the free fermions solutions, at different subsystem sizes. 
The results display nice scaling with $N_S/N$ with collapse of the results for different system sizes, 
typical of the conformal nature of the paramagnetic phase.
In Fig. \ref{fig:AEB-delta-2} and \ref{fig:EAEBE-delta-2} a similar calculation is made at $\Delta=-2$. 
Here, no efficient method is available for calculating the von Neumann mutual information. 
The system is gapped and not conformal, and therefore scaling and collapse with $N_S/N$ are lost.

\section{Conclusions}\label{sec:conclusions}
In this work, we developed efficient calculation methods for the solving of the GEP, 
with the calculation of the maximal R\'enyi divergence in mind, for 1D pure states represented as MPSs.
Our method, adapting the DMRG algorithm and the underlying Lanczos algorithm for the generalized case, 
is efficient and applicable to general GEP systems.
While the maximal divergence can be calculated efficiently, in practice it does not behave as the von Neumann mutual information. 
In the cases studied, the maximal divergence abruptly changes at the phase transition, 
but appears to mirror continuous changes to the mutual information.
This behavior shows that while the maximal divergence cannot be used directly as an alternative for the mutual information,
it can be used to predict phase transitions, and that its behavior is closely tied to that of the mutual information.
Further theoretical review of the maximal divergence could lead to a more refined version, fitting better to the mutual information.

Using this efficient calculation method, the maximal divergence could be studied further in other 
contexts where mutual information plays a key role.
Considering many-body physics, the maximal divergence could be reviewed as a substitution to the mutual
information in characterizing correlation decay \cite{Wolf_2008, Amico_2008} 
and phase transitions \cite{Sukeno_2024, Lang_2020}, as observed here.
In the context of 1+1 conformal theories, the maximal divergence could be investigated as a 
calculable replacement for mutual information for studying universal scaling 
\cite{Alcaraz_2013, Alcaraz_2016} and phase transitions \cite{Alcaraz_2016, Kudler-Flam_2023}, 
substituting existing calculations of R\'enyi mutual information done by path integral replicas \cite{Lashkari_2014, Kudler-Flam_2023}.
In the field of quantum circuits it could be studied in the context of scrambling
\cite{Touil_2020, Hosur_2016}, 
noise and decoherence diagnostics \cite{Niroula_2025, Ahmadi_2024, Fan_2021}, and algorithm benchmarking 
\cite{Illesova_2025, Schumacher_2006}.

This calculation is a general one, useful for many other systems characterized by a GEP. 
As mentioned in Sec. \ref{subsec:GEP}, these include vibrational modes in a lattice \cite{Ashcroft_1976}, 
electromagnetic wave propagation in anisotropic media \cite{Johnson_2001, Joannopoulos_2011, Raghu_2008, Sozuer_1992, Bergmann_1982}, 
and in quantum chemistry with the Roothaan equations \cite{Roothaan_1951, McArdle_2020, Szabo_1996, Ford_1974}.

\section{Acknowledgments}\label{sec:acknowledgments}
We thank S. Scalet for useful discussions.
Our work has been supported by the ISF and the Directorate for Defense Research and Development (DDR\&D)
Grant No. 3427/21, the ISF Grant No. 1113/23, and the US-Israel Binational Science Foundation (BSF) Grants
No. 2020072 and 2024140. 
During this work NF was supported by doctoral fellowships by the Azriely and Milner Foundations, 
then by the Israel Committee for Higher Education Postdoctoral Fellowship.

\appendix

\section{Irrelevance of kernels}\label{sec:Irrelevance-of-Kernel}
The matrices $\rho\equiv\rho_{AB}$, $\sigma\equiv\rho_A\otimes\rho_B$ are positive semi definite 
matrices that could be singular.
While in a general case, randomly chosen positive semidefinite matrices do not necessarily share a kernel space, 
in our case the relation between $\rho$ and $\sigma$ ensures a close 
connection between the kernel spaces of these matrices.

Let $\ket{\psi_A}$ be in the kernel of $\rho_A$, $\rho_A\ket{\psi_A}=0 \to \braket{\psi_A | \rho_A | \psi_A} = 0$.
Then, since
\begin{align}
    &\rho_A = \mathrm{Tr}_B \left[\rho_{AB}\right] = \sum_{i} \braket{\phi^B_i | \rho_{AB} | \phi^B_i},
\end{align}
we have
\begin{align}
    \braket{\psi_A | \rho_A | \psi_A}=\sum_{i} \bra{\psi_A}\braket{\phi^B_i | \rho_{AB} | \phi^B_i}\ket{\psi_A} = 0.
\end{align}
Since $\rho_{AB}$ is positive semidefinite, then for every $i$,  $\braket{\phi^B_i | \rho_{AB} | \phi^B_i} \geq 0$.
To maintain equality to $0$, for every $i$, $\braket{\phi^B_i | \rho_{AB} | \phi^B_i} = 0$.
Therefore, every product state of $\ket{\psi_A}$ with another from subsystem $B$ is in the kernel of $\rho_{AB}$.
In the same way, products of states with states from the kernel of $\rho_B$ are in the kernel of $\rho_{AB}$.
Overall, every kernel state of $\sigma$, will be a kernel state of $\rho$.

With this result in mind, we may now examine the role of kernel vectors of $\sigma$ in the GEP, defined in 
Eq. \eqref{eq:GEP-definition}.
Any vector $\ket{v}\in\mathrm{kernel}(\sigma)\cup\mathrm{kernel}(\rho)$ leads to a trivial identity $0=0$, and thus 
corresponds to an ill-defined generalized eigenvalue, that diverges upon maximization.
We therefore conclude that the kernel subspaces of $\rho,\sigma$ should play no role in the determination of the 
generalized spectrum.

In practice, finite numerical accuracy yield a numerical error, such that the kernel space becomes a
numerical-kernel space: $\left\Vert\sigma\ket{v_0}\right\Vert \gtrsim 0$.
Considering Eq. \eqref{eq:adapted_dmrg}, such numerical kernel vectors pose a difficulty, as they 
dominate the calculation.
Since the theoretical solution should not arise from the kernel, the matrices need to be regularized, to
prevent numerical-kernel solutions.

To better appreciate this, let us consider the example of a bipartite system, displayed in its Schmidt decomposition: 
$\ket{\Phi}=\sum_{i}s_i\ket{v_i^A}\ket{v_i^B}$, where $s_i>0$.
The pure density matrix is given by: $\rho=\sum_{i,j} s_is_j \ket{v_i^A}\ket{v_i^B} \bra{v_j^A}\bra{v_j^B}$.
Then the reduced density matrix are given by: $\rho_{A/B}=\sum_{i}s_i^2 
\ket{v_i^{A/B}}\bra{v_i^{A/B}}$.
Using these forms, we find: 
\begin{align}
    \sigma^{-1/2}\rho\sigma^{-1/2}=\sum_{i,j} \frac{1}{s_j s_i} \ket{v_i^A}\ket{v_i^B} \bra{v_j^A}\bra{v_j^B} .
\end{align}
We thus see that the smallest nonzero singular value $s_0$ will actually dominate, 
and give rise to a maximal generalized eigenvalue $1/s_0^2$.
This shows that numerical kernel vectors can significantly impact the calculation, 
raising the need of regularization, presented in Sec. \ref{subsec:Technical-Implementation-Details}.

\section{Proving equivalence to the generalized Lanczos algorithm}\label{sec:Generalized-Lanczos-eq}

The generalized Lanczos algorithm as appears in Ref. \citep{Parlett_1980}
is adapted from the original Lanczos algorithm, constructing the Krylov
space vectors using:
\begin{align}\label{eq:gen_lanczos_step}
    \beta_{j+1}M\ket{q_{j+1}}&=A\ket{q_{j}}-\alpha_{j}M\ket{q_{j}}-\beta_{j}M\ket{q_{j-1}}\nonumber\\ 
    &\equiv\ket{r_{j+1}}.
\end{align}
Note that in the case of $M=I$, this construction reduces to the
original Lanczos recursion. Based on this, the generalized Lanczos algorithm of Ref. 
\cite{Parlett_1980} is copied here as Alg. \ref{alg:original-generalized-Lanczos}.

\begin{table}[t!]
\begin{minipage}{\linewidth}
\begin{algorithm}[H]
\caption{Generalized Lanczos}
\begin{algorithmic}[1]
    \Require{$A, M, \ket{u_1} \neq 0$}
    \Ensure{Generalized eigenpair $(\theta_j, \ket{s_j})$}
    \State{$\ket{r_1} \gets M\ket{u_1}$}
    \State{$\beta_1 \gets \sqrt{\braket{u_{1}^{*}|r_1}}>0$}
    \For{$j=1,2,...$}
    	\State{Remove non M-orthogonal components of $\ket{u_j}$ if necessary.}
    	\State{$\ket{q_j} \gets \ket{u_j} / \beta_j$}
    	\State{$\ket{\bar{u}_j} \gets A\ket{q_j} - \beta_j \ket{p_{j-1}}$}\label{alg:line:orig-gen-Lanczos-ubar}
    	\State{$\alpha_j \gets \braket{q_j | \bar{u}_j}$}\label{alg:line:orig-gen-Lanczos-alpha}
        \State{$\ket{p_{j}} \gets \ket{r_j}/\beta_j$}\label{alg:line:orig-gen-Lanczos-p}
    	\State{$\ket{r_{j+1}} \gets \ket{\bar{u}_j} - \alpha_j \ket{p_j}$}\label{alg:line:orig-gen-Lanczos-r}
    	\State{Solve $M\ket{u_{j+1}}=\ket{r_{j+1}}$ for $\ket{u_{j+1}}(=\beta_{j+1}\ket{q_{j+1}})$}
    	\State{$\beta_{j+1} \gets \sqrt{\braket{u_{j+1}|r_{j+1}}}$}\label{alg:line:orig-gen-Lanczos-beta}
    	\State{Compute eigenpair $(\theta_j, \ket{s_j})$ of $T_j$ and check for convergence.}
    \EndFor
\end{algorithmic}
\end{algorithm}
\end{minipage}
\caption{\justifying Ref. \citep{Parlett_1980} generalization of the Lanczos algorithm. By using steps 
    \ref{alg:line:orig-gen-Lanczos-ubar}, \ref{alg:line:orig-gen-Lanczos-p} the vectors $\ket{\bar{u}_j}, \ket{p_j}$ 
    can be replaced to form a simpler formulation, showing the recursive nature of $\ket{r_j}$ and included errors 
    in the calculation of $\alpha_j,\beta_j$.}
\label{alg:original-generalized-Lanczos}
\end{table}

This algorithm suffers from numerical instability due to accumulating errors, most notably in $\ket{r_j}, \alpha_j, \beta_j$. 
Using the definitions of Alg. \ref{alg:original-generalized-Lanczos} we can write:
\begin{align}
    \ket{r_{j+1}} &= \ket{\bar{u}_j}-\alpha_j\ket{p_j} = A\ket{q_j}-\alpha_j\ket{p_j}-\beta_j\ket{p_{j-1}}\nonumber\\
        & = A\ket{q_j}-\frac{\alpha_j}{\beta_j}\ket{r_j}-\frac{\beta_j}{\beta_{j-1}}\ket{r_{j-1}},
\end{align}
with the initialization $\ket{r_1}=\beta_1M\ket{q_1}, \ket{r_0}=0$. This is a recursive relation, 
replacing the definition given in Eq. \eqref{eq:gen_lanczos_step}.
Numerical errors in $\ket{q_j}$, resulting from the finite inaccuracy in the M-orthogonalization of 
$\ket{q_j}$, accumulate in each iteration, leading to overall growing numerical error in each 
$\ket{r_j}$. 

Additionally, considering the calculation of $\alpha_j$, $\beta_j$ in Alg. \ref{alg:original-generalized-Lanczos},
the definitions of $\ket{\bar{u}_j}$, $\ket{p_j}$ lead to the following expressions:
\begin{align}
    \alpha_j &= \braket{q_j|A|q_j} - \braket{q_j|r_j},\label{eq:originial-alpha}\\
    \beta_j &= \braket{q_j|A|q_{j-1}} - \alpha_{j-1}\braket{q_j|M|q_{j-1}}\nonumber\\
        &\qquad- \beta_{j-1} \braket{q_j|M|q_{j-2}}.\label{eq:originial-beta}
\end{align}
However, as before, finite inaccuracy in the M-orthogonalization of $\ket{q_j}$ will yield 
additional contributions, resulting in a numerical error.

Below, the steps of Alg. \ref{alg:original-generalized-Lanczos} are used to prove Eqs. \eqref{eq:originial-alpha}
and \eqref{eq:originial-beta}, and M-orthogonality is used to show their equivalence to a more direct formulation. 
Therefore, we adjust the computation of $\alpha_j,\beta_j,\ket{r_j}$, directly utilizing the Krylov vectors $\ket{q_j}$ and the matrices $(A,M)$, to achieve stability and improve accuracy.
This is done at the expense of calculating additional matrix-vector products, as now we need to directly 
calculate such product for $M$.
However, the number of these multiplications can be significantly reduced while increasing memory costs, 
by keeping these products in memory for repeated use.

In our algorithm, the calculation of the intermediate vectors $\ket{r_j}$ 
is done directly by its definition, given by Eq. \eqref{eq:gen_lanczos_step}, shown in Alg. \ref{alg:Generalized-Lanczos}
line \ref{alg:line:gen-Lanczos-r}.
For the calculation of $\alpha_j,\beta_j$ we used a different formulation, as can be seen in Alg. \ref{alg:Generalized-Lanczos}.
We can show equivalency to the definitions of $\alpha_j,\beta_j$ in Alg. \ref{alg:original-generalized-Lanczos} using 
M-orthonormality: $\braket{q_i\left\vert M \right\vert q_j}=\delta_{ij}$.
Starting from the definitions of $\alpha_j, \beta_j$ in lines 
\ref{alg:line:orig-gen-Lanczos-alpha},\ref{alg:line:orig-gen-Lanczos-beta} of Alg. \ref{alg:original-generalized-Lanczos}, 
and using the definitions of $\ket{\bar{u}_j}, \ket{p_{j}}, \ket{r_{j}}$ in lines 
\ref{alg:line:orig-gen-Lanczos-ubar}, \ref{alg:line:orig-gen-Lanczos-p}, \ref{alg:line:orig-gen-Lanczos-r} of Alg. 
\ref{alg:original-generalized-Lanczos}, we show that $\alpha_{j}$ can be calculated using:

\begin{align}
    \alpha_{j}  &\equiv\braket{q_{j}|\bar{u}_{j}}\nonumber\\
        &=\braket{q_{j}|A|q_{j}}-\beta_{j}\braket{q_{j}|p_{j-1}}\nonumber\\
        &=\braket{q_{j}|A|q_{j}}-\frac{\beta_{j}}{\beta_{j-1}}\braket{q_{j}|r_{j-1}}\nonumber\\
        &=\braket{q_{j}|A|q_{j}}-\beta_{j}\cancel{\braket{q_{j}|M|q_{j-1}}}\nonumber\\
        &=\braket{q_{j}|A|q_{j}},
\end{align}
appearing in Alg. \ref{alg:Generalized-Lanczos} line \ref{alg:line:gen-Lanczos-alpha}, and $\beta_j$ can be calculated using:
\begin{align}
    \beta_{j}^2 & = \braket{u_j|r_j} = \braket{u_j|\bar{u}_{j-1}} - \alpha_{j-1}\braket{u_j|p_{j-1}}\nonumber\\
        & = \braket{u_j|A|q_{j-1}} - \beta_{j-1}\braket{u_j|p_{j-2}} - \alpha_{j-1}\braket{u_j|p_{j-1}}\nonumber\\
        & = \beta_j \braket{q_j|A|q_{j-1}} - \beta_{j-1}\beta_j\braket{q_j|p_{j-2}}\nonumber\\
            &\qquad- \alpha_{j-1}\beta_j\braket{q_j|p_{j-1}}\nonumber\\
        & = \beta_j \braket{q_j|A|q_{j-1}} - \beta_{j-1}\beta_j\cancel{\braket{q_j|M|q_{j-2}}}\nonumber\\
        &\qquad - \alpha_{j-1}\beta_j\cancel{\braket{q_j|M|q_{j-1}}}\nonumber\\
        & = \beta_j \braket{q_j|A|q_{j-1}},\nonumber\\  
    \beta_j & =\braket{q_j|A|q_{j-1}}.
\end{align}
appearing in Alg. \ref{alg:Generalized-Lanczos} line \ref{alg:line:gen-Lanczos-beta}.

The Ritz value, which is the norm of the problem residue $\left\Vert \left(A-\theta I\right)\ket{y}\right\Vert $
for an eigenpair $\left(\theta,\ket{y}\right)$, is adapted to the
generalized case as well. Let $Q_{j}$ be the matrix with columns
$\ket{q_{1}}\dots\ket{q_{j}}$, then $AQ_{j}$ is a matrix which columns
are $A\ket{q_{1}}\dots A\ket{q_{j}}$. Similarly, $MQ_{j}$ is a matrix
with columns $M\ket{q_{1}}\dots M\ket{q_{j}}$. Using the tri-diagonal
structure of $T_{j}$, we consider the product $MQ_{j}T_{j}$:
\begin{align*}
    MQ_{j}T_{j}=\left(\begin{array}{ccc}
    | &  & |\\
    M\ket{q_{1}} & \dots & M\ket{q_{j}}\\
    | &  & |
    \end{array}\right)\left(\begin{array}{cccc}
    \alpha_{1} & \beta_{1}\\
    \beta_{1} &  & \ddots\\
     & \ddots &  & \beta_{j-1}\\
     &  & \beta_{j-1} & \alpha_{j}
    \end{array}\right)
\end{align*}
This gives a matrix of which column $i$ is given by:
\begin{align*}
    i=1:\; & \alpha_{1}M\ket{q_{1}}+\beta_{1}M\ket{q_{2}},\\
    1<i<j:\; & \beta_{i-1}M\ket{q_{i-1}}+\alpha_{i}M\ket{q_{i}}+\beta_{i}M\ket{q_{i+1}},\\
    i=j:\; & \beta_{j-1}M\ket{q_{j-1}}+\alpha_{j}M\ket{q_{j}}.
\end{align*}
Now, calculating $AQ_{j}-MQ_{j}T_{j}$ and using Eq. (\ref{eq:gen_lanczos_step})
gives a matrix that is zero in all but the last column, which is given
by: $\ket{r_{j+1}}=A\ket{q_{j}}-\beta_{j-1}M\ket{q_{j}}-\alpha_{j}M\ket{q_{j}}=\beta_{j}M\ket{q_{j+1}}$.
So, similarly to the case of the standard Lanczos (where $M=I$):
\begin{equation}
    \label{eq:Ritz}
    AQ_{j}-MQ_{j}T_{j}=\beta_{j}M\ket{q_{j+1}}\bra{e_{j}},
\end{equation}
where $\ket{e_{j}}$ is the standard basis $j$'th vector: $\ket{e_{j}}=\left(0,0,...,1\right)$.
We can therefore calculate the problem's Ritz value:
\begin{align}
    \label{eq:Ritz-calculation}
    \left\Vert A\ket{y}-\theta M\ket{y}\right\Vert  & =\left\Vert AQ_{j}\ket{s}-\theta MQ_{j}\ket{s}\right\Vert\nonumber\\
     & =\left\Vert \left(AQ_{j}-MQ_{j}T_{j}\right)\ket{s}\right\Vert\nonumber\\
     & =\left\Vert \left(\beta_{j}M\ket{q_{j+1}}\bra{e_{j}}\right)\ket{s}\right\Vert\nonumber\\
     & =\left\Vert \ket{r_{j+1}}\right\Vert \cdot\left|\braket{e_{j}|s}\right|.
\end{align}

Similarly to the standard Lanczos algorithm, this is the measure tested
for convergence, quantifying the error in the calculated eigenpair.
Note that in contrast to the original Lanczos case, where $\left\Vert \ket{q_{j}}\right\Vert =1$,
in the generalized algorithm $\left\Vert M\ket{q_{j}}\right\Vert \neq1$
and therefore $\ket{r_{j+1}}$ must be calculated explicitly.

\section{Correctness and convergence of the generalized Lanczos algorithm}\label{sec:gen-lanczos-correctness}
To prove the correctness of the generalized Lanczos algorithm, appearing in Alg. \ref{alg:Generalized-Lanczos}, 
we consider its final iteration in which the Krylov subspace spanned by $\left\{\ket{q_j}\right\}_{j=1}^n$ is the full space. 
Correctness at this stage, combined with the convergence test of the adapted Ritz value in Eq. \eqref{eq:Ritz-calculation}, 
proves the correctness of the algorithm. 

Let $A, M$ be $(n \times n)$ matrices and let $M$ be invertible. Let $Q=[q_1,q_2,\dots ,q_n]$ be $(n \times n)$ matrix which
columns are M-orthogonal, meaning $\braket{q_i|M|q_j}=\delta_{ij}$, and in matrix notation $Q^\dagger MQ=I_{n \times n}$.
Let $(\lambda, \ket{v})$ be a generalized eigenpair of the matrices $M$, $A$,
\begin{align}
    A\ket{v} =\lambda M\ket{v}.
\end{align}
We can denote: $\ket{v}=Q\ket{x}$:
\begin{align}
    &A\ket{v} =\lambda M\ket{v},\nonumber\\
    &AQ\ket{x} =\lambda MQ\ket{x},\nonumber\\
    &T\ket{x}=Q^\dagger AQ\ket{x} =\lambda Q^\dagger MQ\ket{x} = \lambda\ket{x}.
\end{align}
Every generalized eigenvalue $\lambda$ of $M$, $A$ produces the same eigenvalue of $T\equiv Q^\dagger AQ$. 
The spectrum of $T$ thus coincides with the generalized spectrum of $A$, $M$ exactly, and so the algorithm produces exact results.

Having established the correctness of the algorithm, we move on to provide a bound on the number of iterations required for convergence.
We define the hermitian matrix $C=M^{-1/2}AM^{-1/2}$. 
The constructed Krylov space in the generalized case with $A,M$ using Eq. \ref{eq:gen_lanczos_step}, is equivalent to 
the Krylov space constructed for $C$:
\begin{align}
    &\beta_{j+1}M\ket{q_{j+1}}=A\ket{q_{j}}-\alpha_{j}M\ket{q_{j}}-\beta_{j}M\ket{q_{j-1}},\nonumber\\
    &\beta_{j+1}\ket{\tilde{q}_{j+1}}=M^{-1/2}AM^{-1/2}\ket{\tilde{q}_{j}}-\alpha_{j}\ket{\tilde{q}_{j}}\nonumber\\
    &\qquad-\beta_{j}\ket{\tilde{q}_{j-1}}.
\end{align}
with $\ket{q_j}=M^{-1/2}\ket{\tilde{q}_j}$. 
Therefore the constructed Krylov space is: $\mathcal{K}_m(C, \ket{v})=\mathrm{span}\{\ket{v}, C\ket{v}, \dots, C^{m-1}\ket{v}\}$.
This means that the same convergence bounds of the Lanczos algorithm can be used. 
Therefore, the error in the estimation of the largest generalized eigenvalue, which is the error in the estimation of the largest standard eigenvalue of $C$ is \cite{Golub_2013}:

\begin{equation}
    \lambda_1 - \theta_1 \leq (\lambda_1 - \lambda_n) \left(\frac{\mathrm{tan}(\phi_1)}{c_{m-1}(1+2\rho_1)}\right)^2 ,
\end{equation}
where $(\lambda_i, \ket{\lambda_i})$ are the eigenpairs of $C$, and $\theta_1$ is the estimate of $\lambda_1$, $c_{m-1}(z)$ is the $(m-1)$'th Chebyshev polynomial and:

\begin{equation}
    \rho_1=\frac{\lambda_1-\lambda_2}{\lambda_2-\lambda_n} \quad,\quad \mathrm{cos}(\phi_i)=|\braket{v_1|\lambda_1}|.
\end{equation}

\bibliography{references}

@article{Scalet_2021,
    doi = {10.22331/q-2021-09-14-541},
    url = {https://doi.org/10.22331/q-2021-09-14-541},
    title = {Computable {R}{\'{e}}nyi mutual information: {A}rea laws and correlations},
    author = {Scalet, Samuel O. and Alhambra, {\'{A}}lvaro M. and Styliaris, Georgios and Cirac, J. Ignacio},
    journal = {{Quantum}},
    issn = {2521-327X},
    publisher = {{Verein zur F{\"{o}}rderung des Open Access Publizierens in den Quantenwissenschaften}},
    volume = {5},
    pages = {541},
    month = sep,
    year = {2021}
}

@article{Schollwock_2011,
    title = {The density-matrix renormalization group in the age of matrix product states},
    journal = {Annals of Physics},
    volume = {326},
    number = {1},
    pages = {96-192},
    year = {2011},
    note = {January 2011 Special Issue},
    issn = {0003-4916},
    doi = {https://doi.org/10.1016/j.aop.2010.09.012},
    url = {https://www.sciencedirect.com/science/article/pii/S0003491610001752},
    author = {Ulrich Schollw\"ock},
}

@book{Parlett_1980,
    author = {Parlett, Beresford N.},
    title = {The Symmetric Eigenvalue Problem},
    publisher = {Society for Industrial and Applied Mathematics},
    year = {1998},
    doi = {10.1137/1.9781611971163},
    address = {},
    edition   = {},
    URL = {https://epubs.siam.org/doi/abs/10.1137/1.9781611971163},
    eprint = {https://epubs.siam.org/doi/pdf/10.1137/1.9781611971163}
}

@book{Franchini_2017,
    title={An Introduction to Integrable Techniques for One-Dimensional Quantum Systems},
    ISBN={9783319484877},
    ISSN={1616-6361},
    url={http://dx.doi.org/10.1007/978-3-319-48487-7},
    DOI={10.1007/978-3-319-48487-7},
    journal={Lecture Notes in Physics},
    publisher={Springer International Publishing},
    author={Franchini, Fabio},
    year={2017} 
}

@article{White_1992,
    title = {Density matrix formulation for quantum renormalization groups},
    author = {White, Steven R.},
    journal = {Phys. Rev. Lett.},
    volume = {69},
    issue = {19},
    pages = {2863--2866},
    numpages = {0},
    year = {1992},
    month = {Nov},
    publisher = {American Physical Society},
    doi = {10.1103/PhysRevLett.69.2863},
    url = {https://link.aps.org/doi/10.1103/PhysRevLett.69.2863}
}

@misc{Ghojogh_2023,
    title={Eigenvalue and Generalized Eigenvalue Problems: Tutorial}, 
    author={Benyamin Ghojogh and Fakhri Karray and Mark Crowley},
    year={2023},
    eprint={1903.11240},
    archivePrefix={arXiv},
    primaryClass={stat.ML},
    url={https://arxiv.org/abs/1903.11240}, 
}

@misc{Eisert_2013,
    title={Entanglement and tensor network states}, 
    author={J. Eisert},
    year={2013},
    eprint={1308.3318},
    archivePrefix={arXiv},
    primaryClass={quant-ph},
    url={https://arxiv.org/abs/1308.3318}, 
}

@article{Cirac_2021,
    title = {Matrix product states and projected entangled pair states: Concepts, symmetries, theorems},
    author = {Cirac, J. Ignacio and P\'erez-Garc\'{\i}a, David and Schuch, Norbert and Verstraete, Frank},
    journal = {Rev. Mod. Phys.},
    volume = {93},
    issue = {4},
    pages = {045003},
    numpages = {65},
    year = {2021},
    month = {Dec},
    publisher = {American Physical Society},
    doi = {10.1103/RevModPhys.93.045003},
    url = {https://link.aps.org/doi/10.1103/RevModPhys.93.045003}
}

@article{Wicks_2007,
    title = {Mutual information as a tool for identifying phase transitions in dynamical complex systems with limited data},
    author = {Wicks, R. T. and Chapman, S. C. and Dendy, R. O.},
    journal = {Phys. Rev. E},
    volume = {75},
    issue = {5},
    pages = {051125},
    numpages = {8},
    year = {2007},
    month = {May},
    publisher = {American Physical Society},
    doi = {10.1103/PhysRevE.75.051125},
    url = {https://link.aps.org/doi/10.1103/PhysRevE.75.051125}
}

@misc{swingle_2010,
    title={Mutual information and the structure of entanglement in quantum field theory}, 
    author={Brian Swingle},
    year={2010},
    eprint={1010.4038},
    archivePrefix={arXiv},
    primaryClass={quant-ph},
    url={https://arxiv.org/abs/1010.4038}, 
}

@article{Wolf_2008,
    title = {Area Laws in Quantum Systems: Mutual Information and Correlations},
    author = {Wolf, Michael M. and Verstraete, Frank and Hastings, Matthew B. and Cirac, J. Ignacio},
    journal = {Phys. Rev. Lett.},
    volume = {100},
    issue = {7},
    pages = {070502},
    numpages = {4},
    year = {2008},
    month = {Feb},
    publisher = {American Physical Society},
    doi = {10.1103/PhysRevLett.100.070502},
    url = {https://link.aps.org/doi/10.1103/PhysRevLett.100.070502}
}

@article{Datta_2009,
    author={Datta, Nilanjana},
    journal={IEEE Transactions on Information Theory}, 
    title={Min- and Max-Relative Entropies and a New Entanglement Monotone}, 
    year={2009},
    volume={55},
    number={6},
    pages={2816-2826},
    doi={10.1109/TIT.2009.2018325}
}

@article{tenpy_1,
    title={{Tensor network Python (TeNPy) version 1}},
    author={Johannes Hauschild and Jakob Unfried and Sajant Anand and Bartholomew Andrews and Marcus Bintz and Umberto Borla and Stefan Divic and Markus Drescher and Jan Geiger and Martin Hefel and K\'evin H\'emery and Wilhelm Kadow and Jack Kemp and Nico Kirchner and Vincent S. Liu and Gunnar M\"oller and Daniel Parker and Michael Rader and Anton Romen and Samuel Scalet and Leon Schoonderwoerd and Maximilian Schulz and Tomohiro Soejima and Philipp Thoma and Yantao Wu and Philip Zechmann and Ludwig Zweng and Roger S. K. Mong and Michael P. Zaletel and Frank Pollmann},
    journal={SciPost Phys. Codebases},
    pages={41},
    year={2024},
    publisher={SciPost},
    doi={10.21468/SciPostPhysCodeb.41},
    url={https://scipost.org/10.21468/SciPostPhysCodeb.41},
}

@article{tenpy_2,
    title={{Codebase release 1.0 for TeNPy}},
    author={Johannes Hauschild and Jakob Unfried and Sajant Anand and Bartholomew Andrews and Marcus Bintz and Umberto Borla and Stefan Divic and Markus Drescher and Jan Geiger and Martin Hefel and K\'evin H\'emery and Wilhelm Kadow and Jack Kemp and Nico Kirchner and Vincent S. Liu and Gunnar M\"oller and Daniel Parker and Michael Rader and Anton Romen and Samuel Scalet and Leon Schoonderwoerd and Maximilian Schulz and Tomohiro Soejima and Philipp Thoma and Yantao Wu and Philip Zechmann and Ludwig Zweng and Roger S. K. Mong and Michael P. Zaletel and Frank Pollmann},
    journal={SciPost Phys. Codebases},
    pages={41-r1.0},
    year={2024},
    publisher={SciPost},
    doi={10.21468/SciPostPhysCodeb.41-r1.0},
    url={https://scipost.org/10.21468/SciPostPhysCodeb.41-r1.0},
}

@article{Vidal_2007,
    title = {Classical Simulation of Infinite-Size Quantum Lattice Systems in One Spatial Dimension},
    author = {Vidal, G.},
    journal = {Phys. Rev. Lett.},
    volume = {98},
    issue = {7},
    pages = {070201},
    numpages = {4},
    year = {2007},
    month = {Feb},
    publisher = {American Physical Society},
    doi = {10.1103/PhysRevLett.98.070201},
    url = {https://link.aps.org/doi/10.1103/PhysRevLett.98.070201}
}

@article{Orus_2008,
    title = {Infinite time-evolving block decimation algorithm beyond unitary evolution},
    author = {Or\'us, R. and Vidal, G.},
    journal = {Phys. Rev. B},
    volume = {78},
    issue = {15},
    pages = {155117},
    numpages = {11},
    year = {2008},
    month = {Oct},
    publisher = {American Physical Society},
    doi = {10.1103/PhysRevB.78.155117},
    url = {https://link.aps.org/doi/10.1103/PhysRevB.78.155117}
}

@article{Ostlund_1995,
    title = {Thermodynamic Limit of Density Matrix Renormalization},
    author = {\"Ostlund, Stellan and Rommer, Stefan},
    journal = {Phys. Rev. Lett.},
    volume = {75},
    issue = {19},
    pages = {3537--3540},
    numpages = {0},
    year = {1995},
    month = {Nov},
    publisher = {American Physical Society},
    doi = {10.1103/PhysRevLett.75.3537},
    url = {https://link.aps.org/doi/10.1103/PhysRevLett.75.3537}
}

@book{Golub_2013,
    author = {Golub, Gene H. and Van Loan, Charles F.},
    title = {Matrix Computations - 4th Edition},
    publisher = {Johns Hopkins University Press},
    year = {2013},
    doi = {10.1137/1.9781421407944},
    address = {Philadelphia, PA},
    edition   = {},
    URL = {https://epubs.siam.org/doi/abs/10.1137/1.9781421407944},
    eprint = {https://epubs.siam.org/doi/pdf/10.1137/1.9781421407944}
}

@article{Bergmann_1982,
    author={Bergmann, E. E.},
    journal={The Bell System Technical Journal}, 
    title={Electromagnetic propagation in homogeneous media with Hermitian permeability and permittivity}, 
    year={1982},
    volume={61},
    number={6},
    pages={935-948},
    doi={10.1002/j.1538-7305.1982.tb04324.x}
}

@article{Johnson_2001,
    author = {Steven G. Johnson and J. D. Joannopoulos},
    journal = {Opt. Express},
    number = {3},
    pages = {173--190},
    publisher = {Optica Publishing Group},
    title = {Block-iterative frequency-domain methods for Maxwell's equations in a planewave basis},
    volume = {8},
    month = {Jan},
    year = {2001},
    url = {https://opg.optica.org/oe/abstract.cfm?URI=oe-8-3-173},
    doi = {10.1364/OE.8.000173},
}

@article{Sozuer_1992,
    title = {Photonic bands: Convergence problems with the plane-wave method},
    author = {S\"oz\"uer, H. S. and Haus, J. W. and Inguva, R.},
    journal = {Phys. Rev. B},
    volume = {45},
    issue = {24},
    pages = {13962--13972},
    numpages = {0},
    year = {1992},
    month = {Jun},
    publisher = {American Physical Society},
    doi = {10.1103/PhysRevB.45.13962},
    url = {https://link.aps.org/doi/10.1103/PhysRevB.45.13962}
}

@book{Joannopoulos_2011,
    title={Photonic Crystals: Molding the Flow of Light - Second Edition},
    author={Joannopoulos, J.D. and Johnson, S.G. and Winn, J.N. and Meade, R.D.},
    isbn={9781400828241},
    lccn={2007061025},
    url={https://books.google.co.il/books?id=owhE36qiTP8C},
    year={2011},
    publisher={Princeton University Press}
}

@article{Raghu_2008,
    title = {Analogs of quantum-Hall-effect edge states in photonic crystals},
    author = {Raghu, S. and Haldane, F. D. M.},
    journal = {Phys. Rev. A},
    volume = {78},
    issue = {3},
    pages = {033834},
    numpages = {21},
    year = {2008},
    month = {Sep},
    publisher = {American Physical Society},
    doi = {10.1103/PhysRevA.78.033834},
    url = {https://link.aps.org/doi/10.1103/PhysRevA.78.033834}
}

@book{Born_1996,
    author = {Born, Max and Huang, Kun},
    title = {Dynamical Theory Of Crystal Lattices},
    publisher = {Oxford University Press},
    year = {1996},
    month = {08},
    isbn = {9780192670083},
    doi = {10.1093/oso/9780192670083.001.0001},
    url = {https://doi.org/10.1093/oso/9780192670083.001.0001},
}

@article{Roothaan_1951,
    title = {New Developments in Molecular Orbital Theory},
    author = {Roothaan, C. C. J.},
    journal = {Rev. Mod. Phys.},
    volume = {23},
    issue = {2},
    pages = {69--89},
    numpages = {0},
    year = {1951},
    month = {Apr},
    publisher = {American Physical Society},
    doi = {10.1103/RevModPhys.23.69},
    url = {https://link.aps.org/doi/10.1103/RevModPhys.23.69}
}

@article{McArdle_2020,
    title = {Quantum computational chemistry},
    author = {McArdle, Sam and Endo, Suguru and Aspuru-Guzik, Al\'an and Benjamin, Simon C. and Yuan, Xiao},
    journal = {Rev. Mod. Phys.},
    volume = {92},
    issue = {1},
    pages = {015003},
    numpages = {51},
    year = {2020},
    month = {Mar},
    publisher = {American Physical Society},
    doi = {10.1103/RevModPhys.92.015003},
    url = {https://link.aps.org/doi/10.1103/RevModPhys.92.015003}
}

@book{Szabo_1996,
    title={Modern Quantum Chemistry: Introduction to Advanced Electronic Structure Theory},
    author={Szabo, A. and Ostlund, N.S.},
    isbn={9780486691862},
    lccn={lc96010775},
    series={Dover Books on Chemistry},
    url={https://books.google.co.il/books?id=6mV9gYzEkgIC},
    year={1996},
    publisher={Dover Publications}
}

@article{Ford_1974,
    title = {The generalized eigenvalue problem in quantum chemistry},
    journal = {Computer Physics Communications},
    volume = {8},
    number = {5},
    pages = {337-348},
    year = {1974},
    issn = {0010-4655},
    doi = {https://doi.org/10.1016/0010-4655(74)90011-3},
    url = {https://www.sciencedirect.com/science/article/pii/0010465574900113},
    author = {Brian Ford and George Hall}
}

@book{Wilde_2013, 
    place={Cambridge}, 
    title={Quantum Information Theory}, 
    publisher={Cambridge University Press}, 
    author={Wilde, Mark M.}, 
    year={2013}
}

@article{Alcaraz_2013,
    title = {Universal Behavior of the Shannon Mutual Information of Critical Quantum Chains},
    author = {Alcaraz, F. C. and Rajabpour, M. A.},
    journal = {Phys. Rev. Lett.},
    volume = {111},
    issue = {1},
    pages = {017201},
    numpages = {5},
    year = {2013},
    month = {Jul},
    publisher = {American Physical Society},
    doi = {10.1103/PhysRevLett.111.017201},
    url = {https://link.aps.org/doi/10.1103/PhysRevLett.111.017201}
}

@article{Alcaraz_2016,
    title = {Universal behavior of the Shannon mutual information in nonintegrable self-dual quantum chains},
    author = {Alcaraz, F. C.},
    journal = {Phys. Rev. B},
    volume = {94},
    issue = {11},
    pages = {115116},
    numpages = {11},
    year = {2016},
    month = {Sep},
    publisher = {American Physical Society},
    doi = {10.1103/PhysRevB.94.115116},
    url = {https://link.aps.org/doi/10.1103/PhysRevB.94.115116}
}

@article{Lashkari_2014,
    title = {Relative Entropies in Conformal Field Theory},
    author = {Lashkari, Nima},
    journal = {Phys. Rev. Lett.},
    volume = {113},
    issue = {5},
    pages = {051602},
    numpages = {4},
    year = {2014},
    month = {Jul},
    publisher = {American Physical Society},
    doi = {10.1103/PhysRevLett.113.051602},
    url = {https://link.aps.org/doi/10.1103/PhysRevLett.113.051602}
}

@article{Kudler-Flam_2023,
    title = {R\'enyi Mutual Information in Quantum Field Theory},
    author = {Kudler-Flam, Jonah},
    journal = {Phys. Rev. Lett.},
    volume = {130},
    issue = {2},
    pages = {021603},
    numpages = {6},
    year = {2023},
    month = {Jan},
    publisher = {American Physical Society},
    doi = {10.1103/PhysRevLett.130.021603},
    url = {https://link.aps.org/doi/10.1103/PhysRevLett.130.021603}
}

@article{Touil_2020,
   title={Quantum scrambling and the growth of mutual information},
   volume={5},
   ISSN={2058-9565},
   url={http://dx.doi.org/10.1088/2058-9565/ab8ebb},
   DOI={10.1088/2058-9565/ab8ebb},
   number={3},
   journal={Quantum Science and Technology},
   publisher={IOP Publishing},
   author={Touil, Akram and Deffner, Sebastian},
   year={2020},
   month=may, pages={035005}
}

@article{Hosur_2016,
   title={Chaos in quantum channels},
   volume={2016},
   ISSN={1029-8479},
   url={http://dx.doi.org/10.1007/JHEP02(2016)004},
   DOI={10.1007/jhep02(2016)004},
   number={2},
   journal={Journal of High Energy Physics},
   publisher={Springer Science and Business Media LLC},
   author={Hosur, Pavan and Qi, Xiao-Liang and Roberts, Daniel A. and Yoshida, Beni},
   year={2016},
   month=feb 
}

@article{Niroula_2025,
    title = {Error mitigation thresholds in noisy random quantum circuits},
    author = {Niroula, Pradeep and Gopalakrishnan, Sarang and Gullans, Michael J.},
    journal = {Phys. Rev. B},
    volume = {112},
    issue = {2},
    pages = {024206},
    numpages = {11},
    year = {2025},
    month = {Jul},
    publisher = {American Physical Society},
    doi = {10.1103/qsmz-9kkh},
    url = {https://link.aps.org/doi/10.1103/qsmz-9kkh}
}

@misc{Ahmadi_2024,
    title={Mutual information fluctuations and non-stabilizerness in random circuits}, 
    author={Arash Ahmadi and Jonas Helsen and Cagan Karaca and Eliska Greplova},
    year={2024},
    eprint={2408.03831},
    archivePrefix={arXiv},
    primaryClass={quant-ph},
    url={https://arxiv.org/abs/2408.03831}, 
}

@article{Fan_2021,
    title={Self-organized error correction in random unitary circuits with measurement},
    volume={103},
    ISSN={2469-9969},
    url={http://dx.doi.org/10.1103/PhysRevB.103.174309},
    DOI={10.1103/physrevb.103.174309},
    number={17},
    journal={Physical Review B},
    publisher={American Physical Society (APS)},
    author={Fan, Ruihua and Vijay, Sagar and Vishwanath, Ashvin and You, Yi-Zhuang},
    year={2021},
    month=may 
}

@misc{Illesova_2025,
    title={QMetric: Benchmarking Quantum Neural Networks Across Circuits, Features, and Training Dimensions}, 
    author={Silvie Ill\'esov\'a and Tomasz Rybotycki and Martin Beseda},
    year={2025},
    eprint={2506.23765},
    archivePrefix={arXiv},
    primaryClass={quant-ph},
    url={https://arxiv.org/abs/2506.23765}, 
}

@article{Schumacher_2006,
    title = {Quantum mutual information and the one-time pad},
    author = {Schumacher, Benjamin and Westmoreland, Michael D.},
    journal = {Phys. Rev. A},
    volume = {74},
    issue = {4},
    pages = {042305},
    numpages = {4},
    year = {2006},
    month = {Oct},
    publisher = {American Physical Society},
    doi = {10.1103/PhysRevA.74.042305},
    url = {https://link.aps.org/doi/10.1103/PhysRevA.74.042305}
}

@article{Amico_2008,
    title = {Entanglement in many-body systems},
    author = {Amico, Luigi and Fazio, Rosario and Osterloh, Andreas and Vedral, Vlatko},
    journal = {Rev. Mod. Phys.},
    volume = {80},
    issue = {2},
    pages = {517--576},
    numpages = {0},
    year = {2008},
    month = {May},
    publisher = {American Physical Society},
    doi = {10.1103/RevModPhys.80.517},
    url = {https://link.aps.org/doi/10.1103/RevModPhys.80.517}
}

@article{Sukeno_2024,
    title = {Bulk and boundary entanglement transitions in the projective gauge-Higgs model},
    author = {Sukeno, Hiroki and Ikeda, Kazuki and Wei, Tzu-Chieh},
    journal = {Phys. Rev. B},
    volume = {110},
    issue = {24},
    pages = {245102},
    numpages = {21},
    year = {2024},
    month = {Dec},
    publisher = {American Physical Society},
    doi = {10.1103/PhysRevB.110.245102},
    url = {https://link.aps.org/doi/10.1103/PhysRevB.110.245102}
}

@article{Lang_2020,
    title = {Entanglement transition in the projective transverse field Ising model},
    author = {Lang, Nicolai and B\"uchler, Hans Peter},
    journal = {Phys. Rev. B},
    volume = {102},
    issue = {9},
    pages = {094204},
    numpages = {13},
    year = {2020},
    month = {Sep},
    publisher = {American Physical Society},
    doi = {10.1103/PhysRevB.102.094204},
    url = {https://link.aps.org/doi/10.1103/PhysRevB.102.094204}
}

@article{Groisman_2005,
    title = {Quantum, classical, and total amount of correlations in a quantum state},
    author = {Groisman, Berry and Popescu, Sandu and Winter, Andreas},
    journal = {Phys. Rev. A},
    volume = {72},
    issue = {3},
    pages = {032317},
    numpages = {11},
    year = {2005},
    month = {Sep},
    publisher = {American Physical Society},
    doi = {10.1103/PhysRevA.72.032317},
    url = {https://link.aps.org/doi/10.1103/PhysRevA.72.032317}
}

@book{Ashcroft_1976,
    author = {Ashcroft, Neil W. and Mermin, N. David},
    address = {New York, Fort Worth},
    isbn = {0030839939},
    booktitle = {Solid state physics},
    language = {english},
    publisher = {Holt, Rinehart and Winston; Saunders},
    title = {Solid state physics },
    year = {1976},
    }

@article{Kormos_2017,
    doi = {10.1088/1751-8121/aa70f6},
    url = {https://doi.org/10.1088/1751-8121/aa70f6},
    year = {2017},
    month = {jun},
    publisher = {IOP Publishing},
    volume = {50},
    number = {26},
    pages = {264005},
    author = {Kormos, M\'arton and Zimbor\'as, Zolt\'an},
    title = {Temperature driven quenches in the Ising model: appearance of negative R\'enyi mutual information},
    journal = {Journal of Physics A: Mathematical and Theoretical},
}

@article{Renner_2008,
    author = {Renner, Renato},
    title = {Security of Quantum Key Distribution},
    journal = {International Journal of Quantum Information},
    volume = {06},
    number = {01},
    pages = {1-127},
    year = {2008},
    doi = {10.1142/S0219749908003256},
    URL = {https://doi.org/10.1142/S0219749908003256},
}

@misc{Khatri_2024,
    title={Principles of Quantum Communication Theory: A Modern Approach}, 
    author={Sumeet Khatri and Mark M. Wilde},
    year={2024},
    eprint={2011.04672},
    archivePrefix={arXiv},
    primaryClass={quant-ph},
    url={https://arxiv.org/abs/2011.04672}, 
}

@book{Tomamichel_2016,
    title={Quantum Information Processing with Finite Resources},
    ISBN={9783319218915},
    ISSN={2197-1765},
    url={http://dx.doi.org/10.1007/978-3-319-21891-5},
    DOI={10.1007/978-3-319-21891-5},
    journal={SpringerBriefs in Mathematical Physics},
    publisher={Springer International Publishing},
    author={Tomamichel, Marco},
    year={2016} 
}

@article{Alcaraz_2014,
    title = {Universal behavior of the Shannon and R\'enyi mutual information of quantum critical chains},
    author = {Alcaraz, F. C. and Rajabpour, M. A.},
    journal = {Phys. Rev. B},
    volume = {90},
    issue = {7},
    pages = {075132},
    numpages = {10},
    year = {2014},
    month = {Aug},
    publisher = {American Physical Society},
    doi = {10.1103/PhysRevB.90.075132},
    url = {https://link.aps.org/doi/10.1103/PhysRevB.90.075132}
}

@article{Jean_Marie_2014,
    title = {Shannon and R\'enyi mutual information in quantum critical spin chains},
    author = {St\'ephan, Jean-Marie},
    journal = {Phys. Rev. B},
    volume = {90},
    issue = {4},
    pages = {045424},
    numpages = {18},
    year = {2014},
    month = {Jul},
    publisher = {American Physical Society},
    doi = {10.1103/PhysRevB.90.045424},
    url = {https://link.aps.org/doi/10.1103/PhysRevB.90.045424}
}

@article{Singh_2011,
    title = {Finite-Temperature Critical Behavior of Mutual Information},
    author = {Singh, Rajiv R. P. and Hastings, Matthew B. and Kallin, Ann B. and Melko, Roger G.},
    journal = {Phys. Rev. Lett.},
    volume = {106},
    issue = {13},
    pages = {135701},
    numpages = {4},
    year = {2011},
    month = {Mar},
    publisher = {American Physical Society},
    doi = {10.1103/PhysRevLett.106.135701},
    url = {https://link.aps.org/doi/10.1103/PhysRevLett.106.135701}
}

@article{Banuls_2017,
    title = {Dynamics of quantum information in many-body localized systems},
    author = {Ba\~nuls, M. C. and Yao, N. Y. and Choi, S. and Lukin, M. D. and Cirac, J. I.},
    journal = {Phys. Rev. B},
    volume = {96},
    issue = {17},
    pages = {174201},
    numpages = {9},
    year = {2017},
    month = {Nov},
    publisher = {American Physical Society},
    doi = {10.1103/PhysRevB.96.174201},
    url = {https://link.aps.org/doi/10.1103/PhysRevB.96.174201}
}

@article{Pasquale_2004,
    doi = {10.1088/1742-5468/2004/06/P06002},
    url = {https://doi.org/10.1088/1742-5468/2004/06/P06002},
    year = {2004},
    month = {jun},
    publisher = {},
    volume = {2004},
    number = {06},
    pages = {P06002},
    author = {Pasquale Calabrese and John Cardy},
    title = {Entanglement entropy and quantum field theory},
    journal = {Journal of Statistical Mechanics: Theory and Experiment},
}

@article{Calabrese_2009,
    doi = {10.1088/1751-8113/42/50/504005},
    url = {https://doi.org/10.1088/1751-8113/42/50/504005},
    year = {2009},
    month = {dec},
    publisher = {},
    volume = {42},
    number = {50},
    pages = {504005},
    author = {Calabrese, Pasquale and Cardy, John},
    title = {Entanglement entropy and conformal field theory},
    journal = {Journal of Physics A: Mathematical and Theoretical},
}

@article{Bernigau_2015,
    doi = {10.1088/1742-5468/2015/02/P02008},
    url = {https://doi.org/10.1088/1742-5468/2015/02/P02008},
    year = {2015},
    month = {feb},
    publisher = {IOP Publishing and SISSA},
    volume = {2015},
    number = {2},
    pages = {P02008},
    author = {Bernigau, H and Kastoryano, M J and Eisert, J},
    title = {Mutual information area laws for thermal free fermions},
    journal = {Journal of Statistical Mechanics: Theory and Experiment},
}

@article{Pirvu_2010,
    doi = {10.1088/1367-2630/12/2/025012},
    url = {https://doi.org/10.1088/1367-2630/12/2/025012},
    year = {2010},
    month = {feb},
    publisher = {},
    volume = {12},
    number = {2},
    pages = {025012},
    author = {Pirvu, B and Murg, V and Cirac, J I and Verstraete, F},
    title = {Matrix product operator representations},
    journal = {New Journal of Physics},
}

@article{Cirac_2010,
    title = {Infinite matrix product states, conformal field theory, and the Haldane-Shastry model},
    author = {Cirac, J. Ignacio and Sierra, Germ\'an},
    journal = {Phys. Rev. B},
    volume = {81},
    issue = {10},
    pages = {104431},
    numpages = {4},
    year = {2010},
    month = {Mar},
    publisher = {American Physical Society},
    doi = {10.1103/PhysRevB.81.104431},
    url = {https://link.aps.org/doi/10.1103/PhysRevB.81.104431}
}

@article{Hastings_2010,
    title = {Measuring Renyi Entanglement Entropy in Quantum Monte Carlo Simulations},
    author = {Hastings, Matthew B. and Gonz\'alez, Iv\'an and Kallin, Ann B. and Melko, Roger G.},
    journal = {Phys. Rev. Lett.},
    volume = {104},
    issue = {15},
    pages = {157201},
    numpages = {4},
    year = {2010},
    month = {Apr},
    publisher = {American Physical Society},
    doi = {10.1103/PhysRevLett.104.157201},
    url = {https://link.aps.org/doi/10.1103/PhysRevLett.104.157201}
}

@article{Humeniuk_2012,
    title = {Quantum Monte Carlo calculation of entanglement R\'enyi entropies for generic quantum systems},
    author = {Humeniuk, Stephan and Roscilde, Tommaso},
    journal = {Phys. Rev. B},
    volume = {86},
    issue = {23},
    pages = {235116},
    numpages = {8},
    year = {2012},
    month = {Dec},
    publisher = {American Physical Society},
    doi = {10.1103/PhysRevB.86.235116},
    url = {https://link.aps.org/doi/10.1103/PhysRevB.86.235116}
}

@article{Grover_2013,
    title = {Entanglement of Interacting Fermions in Quantum Monte Carlo Calculations},
    author = {Grover, Tarun},
    journal = {Phys. Rev. Lett.},
    volume = {111},
    issue = {13},
    pages = {130402},
    numpages = {5},
    year = {2013},
    month = {Sep},
    publisher = {American Physical Society},
    doi = {10.1103/PhysRevLett.111.130402},
    url = {https://link.aps.org/doi/10.1103/PhysRevLett.111.130402}
}

@article{Polizzi_2009,
    title = {Density-matrix-based algorithm for solving eigenvalue problems},
    author = {Polizzi, Eric},
    journal = {Phys. Rev. B},
    volume = {79},
    issue = {11},
    pages = {115112},
    numpages = {6},
    year = {2009},
    month = {Mar},
    publisher = {American Physical Society},
    doi = {10.1103/PhysRevB.79.115112},
    url = {https://link.aps.org/doi/10.1103/PhysRevB.79.115112}
}

@article{Verstraete_2004,
    title = {Density Matrix Renormalization Group and Periodic Boundary Conditions: A Quantum Information Perspective},
    author = {Verstraete, F. and Porras, D. and Cirac, J. I.},
    journal = {Phys. Rev. Lett.},
    volume = {93},
    issue = {22},
    pages = {227205},
    numpages = {4},
    year = {2004},
    month = {Nov},
    publisher = {American Physical Society},
    doi = {10.1103/PhysRevLett.93.227205},
    url = {https://link.aps.org/doi/10.1103/PhysRevLett.93.227205}
}

@article{Peschel_2003,
    doi = {10.1088/0305-4470/36/14/101},
    url = {https://doi.org/10.1088/0305-4470/36/14/101},
    year = {2003},
    month = {mar},
    publisher = {},
    volume = {36},
    number = {14},
    pages = {L205},
    author = {Ingo Peschel},
    title = {Calculation of reduced density matrices from correlation functions},
    journal = {Journal of Physics A: Mathematical and General},
}

@misc{klich_2002,
      title={Full Counting Statistics: An elementary derivation of Levitov's formula}, 
      author={I. Klich},
      year={2002},
      eprint={cond-mat/0209642},
      archivePrefix={arXiv},
      primaryClass={cond-mat.mes-hall},
      url={https://arxiv.org/abs/cond-mat/0209642}, 
}

@article{Hestenes_1952,
    title={Methods of conjugate gradients for solving linear systems},
    author={Magnus R. Hestenes and Eduard Stiefel},
    journal={Journal of research of the National Bureau of Standards},
    year={1952},
    volume={49},
    pages={409-435},
    url={https://api.semanticscholar.org/CorpusID:2207234}
}

@article{Eisert_2010,
  title = {Colloquium: Area laws for the entanglement entropy},
  author = {Eisert, J. and Cramer, M. and Plenio, M. B.},
  journal = {Rev. Mod. Phys.},
  volume = {82},
  issue = {1},
  pages = {277--306},
  numpages = {0},
  year = {2010},
  month = {Feb},
  publisher = {American Physical Society},
  doi = {10.1103/RevModPhys.82.277},
  url = {https://link.aps.org/doi/10.1103/RevModPhys.82.277}
}

@misc{Levin_2025,
    author = {Uri Levin},
    title = {Maximal Divergence},
    type = {software},
    note = {\url{https://github.com/uri-levin/MaximalDivergence}},
    year = {2025}
}

\end{document}